\gdef\isHP{T}
\magnification\magstep1

\newdimen\papwidth
\newdimen\papheight
\newskip\beforesectionskipamount  
\newskip\sectionskipamount 
\def\sectionskip{\vskip\sectionskipamount}
\def\beforesectionskip{\vskip\beforesectionskipamount}
\papwidth=16truecm
\papheight=22truecm
\voffset=0.4truecm
\hoffset=0.4truecm
\hsize=\papwidth
\vsize=\papheight
\nopagenumbers
\headline={\ifnum\pageno>1 {\hss\tenrm-\ \folio\ -\hss} \else
{\hfill}\fi}
\newdimen\texpscorrection
\texpscorrection=0.15truecm 

\def\sectionsize{\twelvepoint}
\def\sectiontype{\bf}
\def\subsectionsize{}
\def\subsectiontype{\bf}
\def\em{\sl}

\font\twelverm=ptmr	at	12pt
\font\tenrm=ptmr	at	10pt
\font\eightrm=ptmr	at	8pt
\font\sevenrm=ptmr	at	7pt
\font\sixrm=ptmr	at	6pt
\font\fiverm=ptmr	at	5pt

\font\twelvebf=ptmb	at	12pt
\font\tenbf=ptmb	at	10pt
\font\eightbf=ptmb	at	8pt
\font\sevenbf=ptmb	at	7pt
\font\sixbf=ptmb	at	6pt
\font\fivebf=ptmb	at	5pt

\font\twelveit=ptmri	at	12pt
\font\tenit=ptmri	at	10pt
\font\eightit=ptmri	at	8pt
\font\sevenit=ptmri	at	7pt
\font\sixit=ptmri	at	6pt
\font\fiveit=ptmri	at	5pt

\font\twelvesl=ptmro	at	12pt
\font\tensl=ptmro	at	10pt
\font\eightsl=ptmro	at	8pt
\font\sevensl=ptmro	at	7pt
\font\sixsl=ptmro	at	6pt
\font\fivesl=ptmro	at	5pt

\font\twelvei=cmmi12
\font\teni=cmmi10
\font\eighti=cmmi8
\font\seveni=cmmi7
\font\sixi=cmmi6
\font\fivei=cmmi5

\font\twelvesy=cmsy10	at	12pt
\font\tensy=cmsy10
\font\eightsy=cmsy8
\font\sevensy=cmsy7
\font\sixsy=cmsy6
\font\fivesy=cmsy5
\newfam\truecmsy
\font\twelvetruecmsy=cmsy10	at	12pt
\font\tentruecmsy=cmsy10
\font\eighttruecmsy=cmsy8
\font\seventruecmsy=cmsy7
\font\sixtruecmsy=cmsy6
\font\fivetruecmsy=cmsy5

\newfam\truecmr
\font\twelvetruecmr=cmr12
\font\tentruecmr=cmr10
\font\eighttruecmr=cmr8
\font\seventruecmr=cmr7
\font\sixtruecmr=cmr6
\font\fivetruecmr=cmr5

\font\twelvebf=cmbx12
\font\tenbf=cmbx10
\font\eightbf=cmbx8
\font\sevenbf=cmbx7
\font\sixbf=cmbx6
\font\fivebf=cmbx5

\font\twelvett=cmtt12
\font\tentt=cmtt10
\font\eighttt=cmtt8

\font\twelveex=cmex10	at	12pt
\font\tenex=cmex10

\newfam\msbfam
\font\twelvemsb=msbm10	at	12pt
\font\tenmsb=msbm10
\font\eightmsb=msbm8
\font\sevenmsb=msbm7
\font\sixmsb=msbm6
\font\fivemsb=msbm5

\newfam\scriptfam
\font\twelvescr=eusm10 at 12pt
\font\tenscr=eusm10
\font\eightscr=eusm8
\font\sevenscr=eusm7
\font\sixscr=eusm6
\font\fivescr=eusm5
\def\eightpoint{\def\rm{\fam0\eightrm}%
\textfont0=\eightrm
  \scriptfont0=\sixrm
  \scriptscriptfont0=\fiverm 
\textfont1=\eighti
  \scriptfont1=\sixi
  \scriptscriptfont1=\fivei 
\textfont2=\eightsy
  \scriptfont2=\sixsy
  \scriptscriptfont2=\fivesy 
\textfont3=\tenex
  \scriptfont3=\tenex
  \scriptscriptfont3=\tenex 
\textfont\itfam=\eightit
  \scriptfont\itfam=\sixit
  \scriptscriptfont\itfam=\fiveit 
  \def\it{\fam\itfam\eightit}%
\textfont\slfam=\eightsl
  \scriptfont\slfam=\sixsl
  \scriptscriptfont\slfam=\fivesl 
  \def\sl{\fam\slfam\eightsl}%
\textfont\ttfam=\eighttt
  \def\tt{\fam\ttfam\eighttt}%
\textfont\bffam=\eightbf
  \scriptfont\bffam=\sixbf
  \scriptscriptfont\bffam=\fivebf
  \def\bf{\fam\bffam\eightbf}%
\textfont\scriptfam=\eightscr
  \scriptfont\scriptfam=\sixscr
  \scriptscriptfont\scriptfam=\fivescr
  \def\script{\fam\scriptfam\eightscr}%
\textfont\msbfam=\eightmsb
  \scriptfont\msbfam=\sixmsb
  \scriptscriptfont\msbfam=\fivemsb
  \def\bb{\fam\msbfam\eightmsb}%
\textfont\truecmr=\eighttruecmr
  \scriptfont\truecmr=\sixtruecmr
  \scriptscriptfont\truecmr=\fivetruecmr
  \def\truerm{\fam\truecmr\eighttruecmr}%
\textfont\truecmsy=\eighttruecmsy
  \scriptfont\truecmsy=\sixtruecmsy
  \scriptscriptfont\truecmsy=\fivetruecmsy
\tt \ttglue=.5em plus.25em minus.15em 
\normalbaselineskip=9pt
\setbox\strutbox=\hbox{\vrule height7pt depth2pt width0pt}%
\normalbaselines
\rm
}

\def\tenpoint{\def\rm{\fam0\tenrm}%
\textfont0=\tenrm
  \scriptfont0=\sevenrm
  \scriptscriptfont0=\fiverm 
\textfont1=\teni
  \scriptfont1=\seveni
  \scriptscriptfont1=\fivei 
\textfont2=\tensy
  \scriptfont2=\sevensy
  \scriptscriptfont2=\fivesy 
\textfont3=\tenex
  \scriptfont3=\tenex
  \scriptscriptfont3=\tenex 
\textfont\itfam=\tenit
  \scriptfont\itfam=\sevenit
  \scriptscriptfont\itfam=\fiveit 
  \def\it{\fam\itfam\tenit}%
\textfont\slfam=\tensl
  \scriptfont\slfam=\sevensl
  \scriptscriptfont\slfam=\fivesl 
  \def\sl{\fam\slfam\tensl}%
\textfont\ttfam=\tentt
  \def\tt{\fam\ttfam\tentt}%
\textfont\bffam=\tenbf
  \scriptfont\bffam=\sevenbf
  \scriptscriptfont\bffam=\fivebf
  \def\bf{\fam\bffam\tenbf}%
\textfont\scriptfam=\tenscr
  \scriptfont\scriptfam=\sevenscr
  \scriptscriptfont\scriptfam=\fivescr
  \def\script{\fam\scriptfam\tenscr}%
\textfont\msbfam=\tenmsb
  \scriptfont\msbfam=\sevenmsb
  \scriptscriptfont\msbfam=\fivemsb
  \def\bb{\fam\msbfam\tenmsb}%
\textfont\truecmr=\tentruecmr
  \scriptfont\truecmr=\seventruecmr
  \scriptscriptfont\truecmr=\fivetruecmr
  \def\truerm{\fam\truecmr\tentruecmr}%
\textfont\truecmsy=\tentruecmsy
  \scriptfont\truecmsy=\seventruecmsy
  \scriptscriptfont\truecmsy=\fivetruecmsy
\tt \ttglue=.5em plus.25em minus.15em 
\normalbaselineskip=12pt
\setbox\strutbox=\hbox{\vrule height8.5pt depth3.5pt width0pt}%
\normalbaselines
\rm
}

\def\twelvepoint{\def\rm{\fam0\twelverm}%
\textfont0=\twelverm
  \scriptfont0=\tenrm
  \scriptscriptfont0=\eightrm 
\textfont1=\twelvei
  \scriptfont1=\teni
  \scriptscriptfont1=\eighti 
\textfont2=\twelvesy
  \scriptfont2=\tensy
  \scriptscriptfont2=\eightsy 
\textfont3=\twelveex
  \scriptfont3=\twelveex
  \scriptscriptfont3=\twelveex 
\textfont\itfam=\twelveit
  \scriptfont\itfam=\tenit
  \scriptscriptfont\itfam=\eightit 
  \def\it{\fam\itfam\twelveit}%
\textfont\slfam=\twelvesl
  \scriptfont\slfam=\tensl
  \scriptscriptfont\slfam=\eightsl 
  \def\sl{\fam\slfam\twelvesl}%
\textfont\ttfam=\twelvett
  \def\tt{\fam\ttfam\twelvett}%
\textfont\bffam=\twelvebf
  \scriptfont\bffam=\tenbf
  \scriptscriptfont\bffam=\eightbf
  \def\bf{\fam\bffam\twelvebf}%
\textfont\scriptfam=\twelvescr
  \scriptfont\scriptfam=\tenscr
  \scriptscriptfont\scriptfam=\eightscr
  \def\script{\fam\scriptfam\twelvescr}%
\textfont\msbfam=\twelvemsb
  \scriptfont\msbfam=\tenmsb
  \scriptscriptfont\msbfam=\eightmsb
  \def\bb{\fam\msbfam\twelvemsb}%
\textfont\truecmr=\twelvetruecmr
  \scriptfont\truecmr=\tentruecmr
  \scriptscriptfont\truecmr=\eighttruecmr
  \def\truerm{\fam\truecmr\twelvetruecmr}%
\textfont\truecmsy=\twelvetruecmsy
  \scriptfont\truecmsy=\tentruecmsy
  \scriptscriptfont\truecmsy=\eighttruecmsy
\tt \ttglue=.5em plus.25em minus.15em 
\setbox\strutbox=\hbox{\vrule height7pt depth2pt width0pt}%
\normalbaselineskip=15pt
\normalbaselines
\rm
}
%
\fontdimen16\tensy=2.7pt
\fontdimen13\tensy=4.3pt
\fontdimen17\tensy=2.7pt
\fontdimen14\tensy=4.3pt
\fontdimen18\tensy=4.3pt
\fontdimen16\eightsy=2.7pt
\fontdimen13\eightsy=4.3pt
\fontdimen17\eightsy=2.7pt
\fontdimen14\eightsy=4.3pt
\fontdimen18\eightsy=4.3pt
%
\def\hexnumber#1{\ifcase#1 0\or1\or2\or3\or4\or5\or6\or7\or8\or9\or
 A\or B\or C\or D\or E\or F\fi}
\mathcode`\=="3\hexnumber\truecmr3D
\mathchardef\not="3\hexnumber\truecmsy36
\mathcode`\+="2\hexnumber\truecmr2B
\mathcode`\(="4\hexnumber\truecmr28
\mathcode`\)="5\hexnumber\truecmr29
\mathcode`\!="5\hexnumber\truecmr21
\mathcode`\(="4\hexnumber\truecmr28
\mathcode`\)="5\hexnumber\truecmr29

\def\tilde{\mathaccent"0\hexnumber\truecmr7E }
\def\bar{\mathaccent"0\hexnumber\truecmr16 }

\def\hat{\mathaccent"0\hexnumber\truecmr5E }
\def\dot{\mathaccent"0\hexnumber\truecmr5F }
\def\Phi{\mathchar"0\hexnumber\truecmr08 }
\def\Gamma {\mathchar"0\hexnumber\truecmr00 }
\def\Delta {\mathchar"0\hexnumber\truecmr01 }
\def\Theta {\mathchar"0\hexnumber\truecmr02 }
\def\Lambda{\mathchar"0\hexnumber\truecmr03 }
\def\Xi {\mathchar"0\hexnumber\truecmr04 }
\def\Pi{\mathchar"0\hexnumber\truecmr05 }
\def\Sigma{\mathchar"0\hexnumber\truecmr06 }
\def\Upsilon {\mathchar"0\hexnumber\truecmr07 }
\def\Phi {\mathchar"0\hexnumber\truecmr08 }
\def\Psi {\mathchar"0\hexnumber\truecmr09 }
\def\Omega{\mathchar"0\hexnumber\truecmr0A }
\newcount\EQNcount \EQNcount=1
\newcount\CLAIMcount \CLAIMcount=1
\newcount\SECTIONcount \SECTIONcount=0
\newcount\SUBSECTIONcount \SUBSECTIONcount=1
\def\ifff(#1,#2,#3){\ifundefined{#1#2}%
\expandafter\xdef\csname #1#2\endcsname{#3}\else%
\write16{!!!!!doubly defined #1,#2}\fi}
\def\NEWDEF #1,#2,#3 {\ifff({#1},{#2},{#3})}
\def\actualnumber{\number\SECTIONcount}
\def\EQ(#1){\lmargin(#1)\eqno\tag(#1)}
\def\NR(#1){&\lmargin(#1)\tag(#1)\cr}  
\def\tag(#1){\lmargin(#1)({\rm \actualnumber}.\number\EQNcount)
 \NEWDEF e,#1,(\actualnumber.\number\EQNcount)
\global\advance\EQNcount by 1
}
\def\SECT(#1)#2\par{\lmargin(#1)\SECTION #2\par%
\NEWDEF s,#1,{\actualnumber} %
}
\def\SUBSECT(#1)#2\par{\lmargin(#1)
\SUBSECTION #2\par
\NEWDEF s,#1,{\actualnumber.\number\SUBSECTIONcount}
}
\def\CLAIM #1(#2) #3\par{
\vskip.1in\medbreak\noindent
{\lmargin(#2)\bf #1~\actualnumber.\number\CLAIMcount.} {\sl #3}\par
\NEWDEF c,#2,{#1~\actualnumber.\number\CLAIMcount}
\global\advance\CLAIMcount by 1
\ifdim\lastskip<\medskipamount
\removelastskip\penalty55\medskip\fi}
\def\CLAIMNONR #1(#2) #3\par{
\vskip.1in\medbreak\noindent
{\lmargin(#2)\bf #1.} {\sl #3}\par
\NEWDEF c,#2,{#1}
\global\advance\CLAIMcount by 1
\ifdim\lastskip<\medskipamount
\removelastskip\penalty55\medskip\fi}
\def\SECTION#1\par{\vskip0pt plus.3\vsize\penalty-75
    \vskip0pt plus -.3\vsize
    \global\advance\SECTIONcount by 1
    \beforesectionskip\noindent
{\sectionsize\sectiontype \actualnumber.\ #1}
    \EQNcount=1
    \CLAIMcount=1
    \SUBSECTIONcount=1
    \nobreak\sectionskip\noindent}
\def\SECTIONNONR#1\par{\vskip0pt plus.3\vsize\penalty-75
    \vskip0pt plus -.3\vsize
    \global\advance\SECTIONcount by 1
    \beforesectionskip\noindent
{\sectionsize\sectiontype  #1}
     \EQNcount=1
     \CLAIMcount=1
     \SUBSECTIONcount=1
     \nobreak\sectionskip\noindent}
\def\SUBSECTION#1\par{\vskip0pt plus.2\vsize\penalty-75
    \vskip0pt plus -.2\vsize
    \beforesectionskip\noindent
{\subsectionsize\subsectiontype \actualnumber.\number\SUBSECTIONcount.\ #1}
    \global\advance\SUBSECTIONcount by 1
    \nobreak\sectionskip\noindent}
\def\SUBSECTIONNONR#1\par{\vskip0pt plus.2\vsize\penalty-75
    \vskip0pt plus -.2\vsize
\beforesectionskip\noindent
{\subsectionsize\subsectiontype #1}
    \nobreak\sectionskip\noindent\noindent}
\def\ifundefined#1{\expandafter\ifx\csname#1\endcsname\relax}
\def\equ(#1){\ifundefined{e#1}$\spadesuit$#1\else\csname e#1\endcsname\fi}
\def\clm(#1){\ifundefined{c#1}$\spadesuit$#1\else\csname c#1\endcsname\fi}
\def\sec(#1){\ifundefined{s#1}$\spadesuit$#1
\else Section \csname s#1\endcsname\fi}
\let\endarg=\par
\def\finish{\def\endarg{\par\endgroup}}
\def\start{\endarg\begingroup}

 \def\beginFROM{\start\parskip=0pt\vskip\baselineskip
\def\finish{\def\endarg{\egroup\par\endgroup}}
  \vbox\bgroup\obeylines\eightpoint\em\finish}

\def\ABSTRACT#1\par{
\vskip 1in {\noindent\sectionsize\sectiontype Abstract.} #1 \par}

\def\TODAY{\number\day~\ifcase\month\or January \or February \or March \or
April \or May \or June
\or July \or August \or September \or October \or November \or December \fi
\number\year\timecount=\number\time
\divide\timecount by 60
}
\newcount\timecount
\def\DRAFT{\def\lmargin(##1){\strut\vadjust{\kern-\strutdepth
\vtop to \strutdepth{
\baselineskip\strutdepth\vss\rlap{\kern-1.2 truecm\eightpoint{##1}}}}}
\font\footfont=cmti7
\footline={{\footfont \hfil File:\jobname, \TODAY,  \number\timecount h}}
}
\newbox\strutboxJPE
\setbox\strutboxJPE=\hbox{\strut}
\def\subitem#1#2\par{\vskip\baselineskip\vskip-\ht\strutboxJPE{\item{#1}#2}}
\gdef\strutdepth{\dp\strutbox}
\def\lmargin(#1){}
\def\period{\unskip.\spacefactor3000 { }}
%
%
\newbox\noboxJPE
\newbox\byboxJPE
\newbox\paperboxJPE
\newbox\yrboxJPE
\newbox\jourboxJPE
\newbox\pagesboxJPE
\newbox\volboxJPE
\newbox\preprintboxJPE
\newbox\toappearboxJPE
\newbox\bookboxJPE
\newbox\bybookboxJPE
\newbox\publisherboxJPE
\newbox\inprintboxJPE
\def\refclearJPE{
   \setbox\noboxJPE=\null             \gdef\isnoJPE{F}
   \setbox\byboxJPE=\null             \gdef\isbyJPE{F}
   \setbox\paperboxJPE=\null          \gdef\ispaperJPE{F}
   \setbox\yrboxJPE=\null             \gdef\isyrJPE{F}
   \setbox\jourboxJPE=\null           \gdef\isjourJPE{F}
   \setbox\pagesboxJPE=\null          \gdef\ispagesJPE{F}
   \setbox\volboxJPE=\null            \gdef\isvolJPE{F}
   \setbox\preprintboxJPE=\null       \gdef\ispreprintJPE{F}
   \setbox\toappearboxJPE=\null       \gdef\istoappearJPE{F}
   \setbox\inprintboxJPE=\null        \gdef\isinprintJPE{F}
   \setbox\bookboxJPE=\null           \gdef\isbookJPE{F}  \gdef\isinbookJPE{F}
     
   \setbox\bybookboxJPE=\null         \gdef\isbybookJPE{F}
   \setbox\publisherboxJPE=\null      \gdef\ispublisherJPE{F}
     
}
\def\widestlabel#1{\setbox0=\hbox{#1\enspace}\refindent=\wd0\relax}
\def\ref{\refclearJPE\bgroup}
\def\no   {\egroup\gdef\isnoJPE{T}\setbox\noboxJPE=\hbox\bgroup}
\def\by   {\egroup\gdef\isbyJPE{T}\setbox\byboxJPE=\hbox\bgroup}
\def\paper{\egroup\gdef\ispaperJPE{T}\setbox\paperboxJPE=\hbox\bgroup}
\def\yr{\egroup\gdef\isyrJPE{T}\setbox\yrboxJPE=\hbox\bgroup}
\def\jour{\egroup\gdef\isjourJPE{T}\setbox\jourboxJPE=\hbox\bgroup}
\def\pages{\egroup\gdef\ispagesJPE{T}\setbox\pagesboxJPE=\hbox\bgroup}
\def\vol{\egroup\gdef\isvolJPE{T}\setbox\volboxJPE=\hbox\bgroup\bf}
\def\preprint{\egroup\gdef
\ispreprintJPE{T}\setbox\preprintboxJPE=\hbox\bgroup}
\def\toappear{\egroup\gdef
\istoappearJPE{T}\setbox\toappearboxJPE=\hbox\bgroup}
\def\inprint{\egroup\gdef
\isinprintJPE{T}\setbox\inprintboxJPE=\hbox\bgroup}
\def\book{\egroup\gdef\isbookJPE{T}\setbox\bookboxJPE=\hbox\bgroup\em}
\def\publisher{\egroup\gdef
\ispublisherJPE{T}\setbox\publisherboxJPE=\hbox\bgroup}
\def\inbook{\egroup\gdef\isinbookJPE{T}\setbox\bookboxJPE=\hbox\bgroup\em}
\def\bybook{\egroup\gdef\isbybookJPE{T}\setbox\bybookboxJPE=\hbox\bgroup}
\newdimen\refindent
\refindent=5em
\def\endref{\egroup \sfcode`.=1000
 \if T\isnoJPE
 \hangindent\refindent\hangafter=1
      \noindent\hbox to\refindent{[\unhbox\noboxJPE\unskip]\hss}\ignorespaces
     \else  \noindent    \fi
 \if T\isbyJPE    \unhbox\byboxJPE\unskip: \fi
 \if T\ispaperJPE \unhbox\paperboxJPE\unskip\period \fi
 \if T\isbookJPE {\it\unhbox\bookboxJPE\unskip}\if T\ispublisherJPE, \else.
\fi\fi
 \if T\isinbookJPE In {\it\unhbox\bookboxJPE\unskip}\if T\isbybookJPE,
\else\period \fi\fi
 \if T\isbybookJPE  (\unhbox\bybookboxJPE\unskip)\period \fi
 \if T\ispublisherJPE \unhbox\publisherboxJPE\unskip \if T\isjourJPE, \else\if
T\isyrJPE \  \else\period \fi\fi\fi
 \if T\istoappearJPE (To appear)\period \fi
 \if T\ispreprintJPE Pre\-print\period \fi
 \if T\isjourJPE    \unhbox\jourboxJPE\unskip\ \fi
 \if T\isvolJPE     \unhbox\volboxJPE\unskip\if T\ispagesJPE, \else\ \fi\fi
 \if T\ispagesJPE   \unhbox\pagesboxJPE\unskip\  \fi
 \if T\isyrJPE      (\unhbox\yrboxJPE\unskip)\period \fi
 \if T\isinprintJPE (in print)\period \fi
\filbreak
}
\def\hexnumber#1{\ifcase#1 0\or1\or2\or3\or4\or5\or6\or7\or8\or9\or
 A\or B\or C\or D\or E\or F\fi}
\textfont\msbfam=\tenmsb
\scriptfont\msbfam=\sevenmsb
\scriptscriptfont\msbfam=\fivemsb
\mathchardef\varkappa="0\hexnumber\msbfam7B
\newcount\FIGUREcount \FIGUREcount=0
\newskip\ttglue 
\newdimen\figcenter
\def\figure #1 #2 #3 #4\cr{\null\ifundefined{fig#1}\global
\advance\FIGUREcount by 1\NEWDEF fig,#1,{Fig.~\number\FIGUREcount}\fi
\write16{ FIG \number\FIGUREcount: #1}
{\goodbreak\figcenter=\hsize\relax
\advance\figcenter by -#3truecm
\divide\figcenter by 2
\midinsert\vskip #2truecm\noindent\hskip\figcenter
\includegraphics{#1}\vskip 0.8truecm\noindent \vbox{\eightpoint\noindent
{\bf\fig(#1)}: #4}\endinsert}}
\def\figurewithtex #1 #2 #3 #4 #5\cr{\null\ifundefined{fig#1}\global
\advance\FIGUREcount by 1\NEWDEF fig,#1,{Fig.~\number\FIGUREcount}\fi
\write16{ FIG \number\FIGUREcount: #1}
{\goodbreak\figcenter=\hsize\relax
\advance\figcenter by -#4truecm
\divide\figcenter by 2
\midinsert\vskip #3truecm\noindent\hskip\figcenter
\includegraphics{#1}{\hskip\texpscorrection\input #2 }\vskip 0.8truecm\noindent \vbox{\eightpoint\noindent
{\bf\fig(#1)}: #5}\endinsert}}
\def\fig(#1){\ifundefined{fig#1}\global
\advance\FIGUREcount by 1\NEWDEF fig,#1,{Fig.~\number\FIGUREcount}
\fi
\csname fig#1\endcsname\relax}
\catcode`@=11
\def\footnote#1{\let\@sf\empty 
  \ifhmode\edef\@sf{\spacefactor\the\spacefactor}\/\fi
  #1\@sf\vfootnote{#1}}
\def\vfootnote#1{\insert\footins\bgroup\eightpoint
  \interlinepenalty\interfootnotelinepenalty
  \splittopskip\ht\strutbox 
  \splitmaxdepth\dp\strutbox \floatingpenalty\@MM
  \leftskip\z@skip \rightskip\z@skip \spaceskip\z@skip \xspaceskip\z@skip
  \textindent{#1}\footstrut\futurelet\next\fo@t}
\def\fo@t{\ifcat\bgroup\noexpand\next \let\next\f@@t
  \else\let\next\f@t\fi \next}
\def\f@@t{\bgroup\aftergroup\@foot\let\next}
\def\f@t#1{#1\@foot}
\def\@foot{\strut\egroup}
\def\footstrut{\vbox to\splittopskip{}}
\skip\footins=\bigskipamount 
\count\footins=1000 
\dimen\footins=8in 
\catcode`@=12 

\def\CC{{\script C}}
\def\EE{{\script E}}
\def\HH{{\script H}}
\def\LL{{\script L}}
\def\MM{{\script M}}

\def\OO{{\script O}}

\def\HALF{{\textstyle{1\over 2}}}

\def\real{{\bf R}}

\def\integer{{\bf Z}}
\def\Re{{\rm Re\,}}

\def\PROOF{\medskip\noindent{\bf Proof.\ }}
\def\REMARK{\medskip\noindent{\bf Remark.\ }}
\def\LIKEREMARK#1{\medskip\noindent{\bf #1.\ }}
\tenpoint
\normalbaselineskip=5.25mm
\baselineskip=5.25mm
\parskip=10pt
\beforesectionskipamount=24pt plus8pt minus8pt
\sectionskipamount=3pt plus1pt minus1pt
\def\em{\it}

\if T\isHP
\font\titlefont=ptmb at 14 pt
\font\toplinefont=cmcsc10
\font\pagenumberfont=ptmb at 10pt
\else
\font\titlefont=cmbx10 scaled\magstep2
\font\toplinefont=cmr10
\font\pagenumberfont=cmr10
\let\tenpoint=\rm
\fi
\def\qr{{q,r}}
\def\D{{\truerm D}}
\def\ttilde{{\sim}}
\let\epsilon=\varepsilon
\let\phi=\varphi 
\let\kappa=\varkappa
\def\cc(#1){\tilde #1^{\c}}
\def\pp(#1){\tilde #1^{\perp}}
\def\sign{{\rm sign}}
\def\const{{\rm const.~}}
\def\>{{\rm s}}
\def\<{{\rm c}}
\def\s{{\rm s}}
\def\c{{\rm c}}
\def\FF{{\script F}}
\def\KK{{\script K}}
\def\hide#1{}

\def\dpq{{\Delta(p,q,\tau )}}
\def\dppp{{\Delta(p,p',\tau )}}
\def\gpq{{\Gamma(p,q,\tau )}}
\def\gppp{{\Gamma(p,p',\tau )}}
\let\truett=\tt
\def\ttt{t+t_0}
\def\tttb{(t+t_0)}
\fontdimen3\tentt=2pt\fontdimen4\tentt=2pt
\def\I{|\kern -0.08em |\kern-0.08em|}
\def\tt{\hskip -5truecm\truett
\#\#\#\#\#\#\#\#\#\#\#\#\#\#  }
\def\phicenter(#1){\Phi^{{\rm center}}_{#1}}
\def\phistable(#1){\Phi^{{\rm stable}}_{#1}}
\def\phic{\Phi^{{\rm center}}}
\def\phis{\Phi^{{\rm stable}}}
\def\psic{\Psi^{{2}}}
\def\psis{\Psi^{{3}}}
\def\allx{{x}}
\def\h(#1,#2){h_{#2}^{(#1)}}
\def\CC{{\cal C}}
\def\Hqrn{H_{q,r,\nu}}
\def\HHqrn{{{\cal H}_{q,r,\nu}}}
\def\Hqr{H_{q,r}}
\def\OO{{\cal O}}
\normalbaselineskip=12pt
\baselineskip=12pt
\parskip=0pt
\parindent=22.222pt
\beforesectionskipamount=24pt plus0pt minus6pt
\sectionskipamount=7pt plus3pt minus0pt
\overfullrule=0pt
\hfuzz=2pt
\headline={\ifnum\pageno>1 {\toplinefont Nonlinear Stability}
\hfill{\pagenumberfont\folio}\fi}
{\titlefont{\centerline{Geometric Stability Analysis for Periodic 
Solutions }}}
\vskip 0.5truecm
{\titlefont{\centerline {of the Swift-Hohenberg Equation}}
\vskip 0.5truecm
{\it{\centerline{J.-P. Eckmann${}^{1,2}$, C.E. Wayne${}^3$, and P. Wittwer${}^1$}}
\vskip 0.3truecm
{\eightpoint
\centerline{${}^1$D\'ept.~de Physique Th\'eorique, Universit\'e de Gen\`eve,
CH-1211 Gen\`eve 4, Switzerland}
\centerline{${}^2$Section de Math\'ematiques, Universit\'e de Gen\`eve,
CH-1211 Gen\`eve 4, Switzerland}
\centerline{${}^3$Dept.~of Mathematics, The Pennsylvania State University,
University Park, PA 16802, USA}
}}
\vskip 0.5truecm
{\eightpoint\narrower\baselineskip 11pt
\LIKEREMARK{Abstract}In this paper we describe invariant geometrical
structures in the 
phase space of the Swift-Hohenberg equation in a neighborhood of its
periodic stationary states. We show that in spite of the fact that
these states are only marginally stable ({\it i.e.}, the linearized
problem about these states has continuous spectrum extending all the
way up to zero), there exist finite dimensional invariant manifolds in
the phase space of this equation which determine the long-time behavior
of solutions near these stationary solutions. In particular, using
this point of view, we obtain a new demonstration of Schneider's recent
proof that these states are nonlinearly stable.}
\vfill\eject
\tenpoint
\def\tt{\truett
\#\#\#\#\#\#\#\#\#\#\#\#\#\#  }
\SECTION Introduction

In this paper, we study the non-linear stability of space-periodic,
time-independent solutions of the Swift-Hohenberg equation
$$
\partial_t u \,=\, \bigl (\epsilon ^2 - (1+\partial_x^2)^2\bigr )u -
u^3~.
\EQ(sh)
$$ 
Here, $u(x,t)$ is defined on $\real \times \real^+$ and takes real
values
and $\epsilon \ge 0$ is a small parameter.
The Eq.\equ(sh) has stationary solutions
$u(x,t)=u_{\epsilon , \omega }(x)$ which are of the form
$$
u_{\epsilon ,\omega }(x)\,=\, \sum_{n\in\integer} u_{\epsilon ,\omega
,n} e^ {i \omega  n x} ~.
\EQ(per)
$$
The non-linear stability problem addresses the question of the time
evolution of initial data which are close to $u_{\epsilon ,\omega }$,
and stability in this context means that the solution converges to
$u_{\epsilon ,\omega }$ as $t\to \infty $.
The range of possible values of $\omega $ is given
by $\epsilon ^2 > (1-\omega ^2)^2$ when $\omega $ is close to 1.
To simplify the exposition we shall concentrate on the case $\omega
=1$, and omit henceforth the index $\omega $.

In a very interesting paper, G. Schneider [Sch] has solved this
problem, and the present work relies heavily on his ideas. Our aim is
to simplify somewhat the exposition of [Sch] and to extend the result
by giving a more precise asymptotic analysis, using the description of
the asymptotic behavior in terms of a continuous renormalization
group and invariant manifolds as introduced in [W], see below.

The {\em existence} of solutions of the form Eq.\equ(per) is a
well-established fact, (see {\it e.g.}~[CE]) and we repeat here only those
points of the discussion which are needed in the sequel. The equation
for the stationary solution is $F(u,\epsilon )=0$, where
$$
F(u,\epsilon )\,\equiv\, \bigl (\epsilon  ^2-(1+\partial_x^2)^2\bigr ) u
-u^3~.
\EQ(fu)
$$
The equation $F=0$ has the trivial solution $u=0$, $\epsilon
=0$. Linearizing around this solution, we see that $\D F$ equals
$$
\D F\,=\, -(1+\partial_x^2)^2 \oplus 0~,
$$
acting on some weighted subspace of $L^2(\real)\oplus\real$. The
null space of $\D F$ is spanned by
$$
\{ \cos x, \sin x \} \oplus 0  \quad {\rm and}\quad 0\oplus 1~,
\EQ(basis)
$$
and thus, bifurcation theory suggests the existence of solutions of the
form of Eq.\equ(per), when $\epsilon \ne0$. This is indeed what
happens ({\it cf.}~[CR], [CE]), and the higher frequency terms in
Eq.\equ(per) are generated from the basis Eq.\equ(basis) by the
non-linearity $u^3$. The method clearly extends to similar polynomial
non-linearities. An explicit calculation
shows that $F(u_\epsilon ,\epsilon )=0$ for
$$
u_\epsilon (x)\,=\, \epsilon  {2\over \sqrt{3}} \cos(x) + \epsilon ^2
h_\epsilon (x)~,
\EQ(sepsilon)
$$
and $h_\epsilon (x)=h_\epsilon (x+2\pi)$. Thus, the function
$u_\epsilon $ equals $u_{\epsilon ,1}$ of Eq.\equ(per).
We have broken the translation invariance of the problem by the choice
of $\cos$ in Eq.\equ(sepsilon), instead of, say, $\sin$.

We next pass to the {\em linear stability analysis} of the solution
$u_\epsilon$. This is again a classical subject, initiated by Eckhaus
[E], which we summarize for convenience, see also [CE]. Linearizing
Eq.\equ(sh) around the solution $u_\epsilon $ we are led to study the
operator
$L_\epsilon =\bigl (\epsilon ^2-(1+\partial_x^2)^2\bigr ) -
3u_\epsilon ^2$,
that is,
$$
\bigl (L_\epsilon v\bigr )(x)\,=\,\bigl (\epsilon ^2-3u_\epsilon
^2(x)\bigr ) v(x) -(1+\partial_x^2)^2 v(x)~.
$$
Because $u_\epsilon $ is a $2\pi$ periodic function,
it is most convenient to work in Floquet coordinates ({\it i.e.}, with Bloch
waves).
To fix the notation, we give some details:
Begin by introducing the following
representation for $f\in L^2(\real)$:
$$
\eqalign{
f(x)\,&=\,\int dk e^{-ikx} \hat f (k) \,=\,
\sum_{m\in\integer} \int_{-1/2}^{1/2} d\ell\, e^{-imx} e^{-i\ell x}
\hat f(m+\ell)\cr
\,&=\,
\int_{-1/2}^{1/2} d\ell\, e^{-i\ell x} \tilde f_\ell (x)~, 
}
$$
where
$$
\tilde f_\ell (x)\,=\,\sum_{m\in\integer}  e^{-imx}
\hat f(m+\ell)~.
\EQ(ftild)
$$
\LIKEREMARK{Properties of $\tilde f$}Note first that $\tilde f_\ell$ is
$2\pi$ periodic. 
Furthermore, the definition of $\tilde f_\ell(x)$
can be extended to all $\ell\in\real$ by the definition
$$
\tilde f_{\ell+1}(x)\,=\,e^{-ix}\tilde f_\ell(x)~.
$$
We next observe that if $f$ has a smooth, rapidly decaying Fourier
transform, then $\tilde f_\ell(x)$ will also be a smooth function of
$\ell $ and $x$.
If $f$, $g$ are
in $L^2(\real)$, then it follows from the definition of $\tilde
f_\ell$ that
$$
(fg)_\ell^\sim(x)\,=\,\int_{-1/2}^{1/2}dk\, \tilde f_{\ell-k}(x)\,
\tilde g_k(x)~.
\EQ(convol)
$$
We finally note that if $s$ is a $2\pi$ periodic function, then
$$
\tilde s_\ell (x) \,=\, \delta(\ell) s(x)~.
\EQ(stilde)
$$
It is now easy to see that
$$
\bigl (L_\epsilon v\bigr)_\ell^\ttilde(x)\,=\,
\bigl (\epsilon ^2-(1+(i\ell+\partial_x)^2)^2\bigr ) \tilde v_\ell (x) 
- 3 (u_\epsilon ^2v )_\ell ^\ttilde (x)~.
$$
In the language of condensed matter physics, $\ell$ is the 
quasi-momentum in the ``Brillouin
zone'' $[-\HALF,\HALF]$ and 
$L_\epsilon $ leaves the subspace $\FF_\ell $ of functions with
quasi-momentum $\ell $ invariant.
Using the properties just described, we get 
$$
\bigl (L_\epsilon v\bigr)_\ell^\ttilde(x)\,=\,
\bigl (\epsilon ^2-(1+(i\ell+\partial_x)^2)^2\bigr ) \tilde v_\ell (x) 
- 3 u_\epsilon ^2(x)\cdot \tilde v _\ell  (x)\,\equiv \bigl (L_{\epsilon
,\ell} v_\ell\bigr )(x)~.
\EQ(loperator)
$$
To fix the notation, we repeat the
calculation done by Eckhaus, {\it cf.}~also [CE], [M]. We denote
$c(x)=\cos(x)$,
$s(x)=\sin(x)$. The method of Eckhaus consists in projecting the
eigenvalue problem for $L_{\epsilon,\ell}$ onto the subspace spanned by the
``bifurcating directions'' $c$ and $s$.
Observe that, modulo higher frequency terms, we have
$c^3={3\over 4} c$, $c^2 s = {1\over 4} s$, and therefore the
projection of $L_{\epsilon,\ell} $ onto this subspace is described by
the matrix
$$
\left (\matrix { -4\ell^2 -\ell^4 -2\epsilon
^2+\OO(\epsilon ^4) & -4 i\ell^3\cr
4 i\ell^3 & -4 \ell^2 -\ell^4}\right )+\OO(\epsilon ^4)
\left (
\matrix{
\OO(\ell^2) &\OO(\ell)\cr
\OO(\ell) & \OO(\ell^2)\cr
}\right )~.
$$
The eigenvalues of this matrix are
$$
\eqalign{
\lambda_{\ell,0}^0\,&=\,- \bigl (4+\OO(\epsilon ^2)\bigr ) \ell ^2
+\OO(\ell ^3)~,\cr
\lambda_{\ell,1}^0\,&=\,-2\bigl (\epsilon ^2 +\OO(\epsilon ^4)\bigr )- \bigl (4+\OO(\epsilon ^2)\bigr ) \ell ^2
+\OO(\ell ^3)+\OO( \ell ^4 +\epsilon ^4)~.\cr
}
$$
Thus, the restriction of $L_{\epsilon ,\ell}$ on the subspace spanned
by $c$ and $s$ has its spectrum in the left half-plane.
Note that the corresponding eigenvectors are 
$s + {\OO}(\ell + \epsilon)$
and $c + {\OO}(\ell + \epsilon)$. 
Extending this calculation to the full space, one shows in the same
way [E, CE, M] that
\CLAIM Theorem(Eckhaus) For sufficiently small $\epsilon >0$ the
operators $L_{\epsilon,\ell}$, with $\ell\in [-\HALF, \HALF]$ are
selfadjoint on the Sobolev space $H^4$,
have compact resolvent and a spectrum satisfying
$$
\eqalign{
\lambda_{\ell,0}(\epsilon ^2)\,&=\,\,- \bigl (4+\OO(\epsilon ^2)\bigr ) \ell ^2
+\OO(\ell ^3)\,\equiv\,-c_0(\epsilon ^2) \ell ^2+\OO(\ell^3)~,\cr
\lambda_{\ell,1}(\epsilon ^2)\,&=\,\,-2\bigl (\epsilon ^2
+\OO(\epsilon ^4)\bigr) - \bigl (4+\OO(\epsilon ^2)\bigr ) \ell ^2 
+\OO(\ell ^3)~,\cr
\lambda_{\ell,j}\,&\le\,-(1-j^2)^2+\OO(\epsilon ^2)~, \quad j=2,3,\dots~.\cr
}
\EQ(spec)
$$  

\LIKEREMARK{Notation}Since we mostly concentrate on the branch 0, we
shall abbreviate $\lambda_\ell=\lambda_{\ell,0}(\epsilon ^2)$.
The eigenfunction corresponding to $\lambda_\ell $ is
$$
\phi _{\epsilon,\ell}(x) \,=\,\const \bigl (\,u_\epsilon '(x)+i\ell g_\epsilon (x) +
h_{\epsilon,\ell } (x) \ell^2~\bigr )~,
\EQ(phiell)
$$
where $u_\epsilon $ is the stationary solution, and both $g_\epsilon$
and $ 
h_{\ell,\epsilon }$ are $2\pi$ periodic. If we choose the constant to
normalize the $L^2$ norm of $\phi _{\epsilon,\ell}$ to 1, 
then $\phi _{\epsilon,\ell} =\pi^{-1/2} \sin(x)+\OO(\epsilon+| \ell| )$.

We can now formulate the main question of this paper: Having seen that
the solution $u_\epsilon $ is linearly (marginally) stable, is it true
that this solution is {\em stable} under the non-linear evolution? The
answer will be affirmative. As pointed out by Schneider [Sch], the
result is not obvious, since the leading non-linear term does not have
a sign. Indeed, the non-linear evolution equation for a (small)
perturbation of $u_\epsilon $ is
$$
\partial_t v\,=\,  -(1+\partial_x^2)^2 v + \epsilon^2 v - 3u_\epsilon
^2 v - 3 u_\epsilon v^2 - v^3~,
$$
where we recall that $u_\epsilon $ is of order $\epsilon $, and
approximately equal to
$\OO(\epsilon) \cos(x)$.
Reducing again to quasi-momentum $\ell$,
and using Eq.\equ(stilde),
we get the equation
$$
\partial_t \tilde v_\ell\,=\,  L_{\epsilon ,\ell} \tilde v_\ell 
- 3 u_\epsilon
(v^2)_\ell^\sim - (v^3)_\ell^\sim~,
\EQ(v1)
$$
and it is the term $3 u_\epsilon (v^2)_\ell^\sim$ which does not have a
sign. The saving grace will be the {\em diffusive} behavior suggested by
the spectrum (in particular the branch
$\lambda_\ell $).
At first sight, the non-linearities seem to be too singular for
diffusion to dominate a potential divergence. Indeed, it is well known
that, {\it e.g.}, the equation
$$
\partial_t u\,=\, \partial_x^2 u+ u^3~,
$$ 
has solutions which blow up in finite time [L], and the quadratic term
makes things even worse. The beautiful observation of Schneider[Sch]
is,
however, that the problem
Eq.\equ(v1) is rather of a form
reminiscent of
$$
\partial_t v\,=\,\partial_x^2 v - \partial_x ^2 ( v^2 + v^3)~,
\EQ(mechan)
$$
which is good enough for convergence [CEE, BK, BKL].

In later sections we examine in detail the form of the non-linear
terms in Eq.\equ(v1), but here we explain briefly why these terms are
similar to the non-linear terms in Eq.\equ(mechan).
The derivatives in the non-linearity
have their origin in the symmetries of the problem,
and they are easier to understand in momentum space.
In fact, Eq.\equ(mechan) is a good approximation to Eq.\equ(v1) 
only in the low-momentum (small $\ell$) regime, but this is sufficient
since for $\ell$ outside a neighborhood of $\ell=0$, the stationary
solutions are linearly stable, (and not only 
{\em marginally} stable) and the form of the non-linearity
is unimportant.

To understand the low-momentum behavior
of Eq.\equ(sh), note first that the Swift-Hohenberg
equation Eq.\equ(sh)---and, incidentally, other equations with
coordinate independent right hand side---has a {\em circle of fixed
points} generated by translations.
If we now study the Eq.\equ(v1) at $\ell=0$, this corresponds to
studying the Swift-Hohenberg equation
in the space of functions of period $2\pi$.
In this space, say $L^2([0,2\pi])$,
the linear operator in Eq.\equ(v1) has pure point
spectrum with a simple eigenvalue at 0 and all other eigenvalues real
and strictly negative.
In this case, as Schneider notes, the center manifold theorem can be
applied, and there exists a 1-dimensional center manifold. We also
see immediately that the eigenvector corresponding to the 0
eigenvalue is $\partial_x u_\epsilon$, {\it i.e.}, it is tangent to the
circle of fixed points generated by translations.
In fact, since any fixed point sufficiently close to the origin must
lie in the center manifold,
we see that {\em the center manifold coincides with the 1-dimensional
circle of fixed points}. Thus the non-linearity in the equation, when
restricted to the center manifold, {\em must vanish}.
This shows that the effective non-linearity in
Eq.\equ(v1), when evaluated at $\ell=0$, must vanish and this accounts
for one derivative in Eq.\equ(mechan). More precisely, we see that the
effective non-linearity in
Eq.\equ(v1) is bounded by $\OO(\ell)$, as is the non-linearity in
Eq.\equ(mechan). The second derivative of the non-linearity in
Eq.\equ(mechan) arises because of ``momentum conservation.'' Since 
$\phi_{\epsilon ,\ell}$ is a smooth function of $\ell$, the
linear term in Eq.\equ(phiell) must of the form $i\ell
g_\epsilon $, with $g_\epsilon $ independent of $\ell$. Since the
interaction is local in $x$, one sees upon working out the integrals
that all terms proportional to $\ell$ in the non-linearity cancel
exactly,
see Eq.(A.3). Thus, the low momentum behavior of Eq.\equ(v1)
is as if the non-linearity was differentiated twice---{\it i.e.},
exactly as in Eq.\equ(mechan).

Our main result is that this intuitive argument correctly predicts that
the leading order asymptotics are diffusive, and that furthermore,
the higher order asymptotics are controlled by a sequence of finite
dimensional invariant manifolds. Thus, our approach provides some 
insight into how
finite dimensional geometrical structures can arise from a problem with
continuous spectrum.
\CLAIM Stability Theorem(stab) Fix $n \ge 1$ and 
$\delta > 0$. There exists a Hilbert space,
${\cal H}(n)$, such that if the initial conditions of \equ(v1) lie in
a sufficiently small neighborhood of the origin in ${\cal H}(n)$, then there
exists an $n+1$ dimensional, invariant manifold in the extended phase
space $P(n) = \real^+ \times {\cal H}(n)$ of
\equ(v1), and any sufficiently small solution of this equation which is
not on this manifold approaches it at a rate 
${\OO}(t^{-(n+1-\delta)/2})$. In
particular, if $n=1$, small solutions of \equ(v1) have the asymptotic
form:
$$
v(x,t) \,=\, {{A}\over{\sqrt{t}}} e^{-x^2/4 t} + 
{\OO}({{1}\over{t^{3/4-\delta}}})~.
$$

\REMARK In Sections 2 and 3, we will make clear precisely what the
Hilbert spaces ${\cal H}(n)$ are and what we mean by ``sufficiently small.''

The remainder of the paper is devoted to a proof of the \clm(stab).
\SECTION Formulating the \clm(stab) in terms of scaling variables

In this section, we transform the problem to a rescaled dynamical
system.
In the next section, we will cast the dynamical system thus obtained
into an invariant manifold problem. 

The idea of the proof is to focus on the ``central branch'' of the spectrum,
$\lambda_{\ell} = \lambda_{\ell,0}(\epsilon^2)$, which is only marginally
stable. The relevant part of the spectrum for the long-time asymptotics
is only the part in a small neighborhood of $\ell =0$, a fact we 
exhibit by an appropriate rescaling of the dependent and independent
variables. This rescaling has the disadvantage that it introduces a
singular 
perturbation in the variables corresponding to the ``stable branches''
of the spectrum, $\lambda_{\ell,n}(\epsilon^2)$, $n \ge 1$, because
the corresponding modes decay extremely fast, when rescaled (at least
on a linear level).
However,
invariant manifold theory has long been used to treat singular perturbation
problems, and we are able to use it for that purpose here as well.
In addition, these invariant manifolds will provide us with a geometric
description of the long-time asymptotics of solutions near the stationary
states.

Our method generalizes to other problems of similar spectral nature,
see the example of a cylindrical domain given in [W2].

Henceforth, we fix $\epsilon>0$, and omit it from most
subscripts.
Since $L_\ell=L_{\epsilon ,\ell}$ is self-adjoint, we can define the 
(orthogonal) 
spectral projections
$P_\ell$ and $P_\ell^\perp$, which project onto
the central branch and its complement.
\REMARK We know that for $|\ell|$ sufficiently small, say
$|\ell|<\ell_0/2$, one has 
$$
{\rm spec} ( P_\ell L_\ell P_\ell)\,=\, -c_0(\epsilon^2)
\ell^2+\OO(\ell^3)~,
$$
and that this is the eigenvalue closest to 0 in ${\rm spec}( L_\ell)$.
We continue this projection smoothly to larger $\ell$ even if it
cannot be guaranteed to be a projection onto the highest
eigenvalue. But note that for those values of $\ell$ the spectrum of
$L_\ell$ can
be shown to be strictly bounded away from 0, see, {\it e.g.}, [CE,
page 102]. 

To study the non-linearity, and to show the mechanism leading to the
result which is analogous to Eq.\equ(mechan), we write the Eq.\equ(v1)
in more detail:
$$
\eqalign{
\partial_t \tilde v_\ell(x)\,&=\,  \bigl (L_{\epsilon ,\ell} \tilde
v_\ell\bigr ) (x) -  3 u_\epsilon(x)\hide{ e^{-i\ell x}}
\int_{-1/2}^{1/2}\kern -1em dk\,
\tilde v_{\ell-k}(x) \tilde v_k(x)\cr  & \qquad - \hide{e^{-i\ell x}} 
\int_{-1/2}^{1/2}\kern -1em dk_1\,dk_2\, \tilde v_{\ell-k_1-k_2}(x)
\tilde v_{k_1}(x)\tilde v_{k_2}(x)\cr
\,&\equiv\,\bigl (L_{\ell} \tilde
v_\ell\bigr )(x)- \bigl (F_2(\tilde v)\bigr )_\ell(x)
-\bigl (F_3(\tilde v)\bigr )_\ell(x)
~.\cr
}
\EQ(theequ)
$$
We now decompose the Eq.\equ(theequ) by projecting onto $P_\ell$ and $P_\ell^\perp$.
If $f\in L^2$, we let $\cc(f_\ell)=P_\ell \tilde f_\ell$, and
$\pp(f_\ell)=
P_\ell^\perp \tilde f_\ell$. Similarly, $L_\ell^\c  = P_\ell L_\ell
P_\ell$ and
$L_\ell^\perp= P_\ell^\perp L_\ell P_\ell^\perp$.
Then we get
$$
\eqalign{
\partial_t \cc(v_\ell)(x)\,&=\, L_\ell^\c  \cc(v_\ell) (x) -
\bigl (P_\ell F_2(\tilde v)_\ell\bigr )(x)
- \bigl (P_\ell F_3(\tilde v)_\ell\bigr )(x)~,\cr
}
\EQ(ccv)
$$
and a similar equation for $\pp(v_\ell)$:
$$
\eqalign{
\partial_t \pp(v_\ell)(x)\,&=\, L_\ell^\c  \pp(v_\ell)(x) -
\bigl (P_\ell^\perp F_2(\tilde v)_\ell\bigr )(x)
- \bigl (P_\ell^\perp F_3(\tilde v)_\ell\bigr )(x)~.\cr
}
\EQ(ppv)
$$
We next split the first equation into a piece corresponding to small
$|\ell|$, {\it i.e.}, $|\ell|<\ell_0$ and another corresponding to large
$\ell$. Since we want to construct invariant manifolds, we need some
smoothness in this construction and we choose a smooth cutoff
$\chi$ satisfying
$$
\chi(\ell)\,=\,\cases  { 1, & if $|\ell|\le \ell_0$ ~,\cr
0, & if $|\ell|>2 \ell_0$~,
}
$$
and of course $\ell_0<\HALF$. In fact, we shall choose $\ell_0>0$ so
small that $P_\ell$ is the projection onto the central eigenspace for
all $\ell\in[-\ell_0,\ell_0]$. Let $\phi_\ell$ denote the normalized
eigenvector which spans the range of $P_\ell$ (for $|\ell| <\ell_0$,
and smoothly continued for $\ell$ beyond that value). Then
$\cc(v_\ell)$ can be written as $\cc(v_\ell)= V(\ell) \phi_\ell$,
where it is understood that $V$ is really a function of $v$. We also
let $\Pi_\ell$ denote the operation $\Pi_\ell f_\ell = \langle
\phi_\ell | f_\ell\rangle$, where $\langle \cdot \rangle$ is the
scalar product in $\FF_\ell$. This operation extracts the coefficient
$V$ and therefore Eq.\equ(ccv) can be written as
$$
\eqalign{
\partial_t V(\ell)\,&=\, \lambda_\ell   V(\ell)  -
\Pi_\ell P_\ell F_2(\tilde v)_\ell
- \Pi_\ell P_\ell F_3(\tilde v)_\ell~.\cr
}
\EQ(av)
$$
Defining $V^{<}(\ell)=\chi(\ell) V(\ell) $, and 
$V^{>}(\ell)=(1-\chi(\ell)) V(\ell) $, the Eq.\equ(av) can be rewritten
as
$$
\eqalign{
\partial_t V^{<}(\ell)\,&=\, \lambda_\ell   V^{<}(\ell)-
\bigl (f^\<(V^{<},V^{>},\pp(v))\bigr )(\ell)~,\cr
 \partial_t V^{>}(\ell)\,&=\, \lambda_\ell   V^{>}(\ell)-
\bigl (f^\>(V^{<},V^{>},\pp(v))\bigr )(\ell)~,\cr
}
\EQ(aggv)
$$ 
where
$$
\eqalign{
\bigl ( f^\c(V^{<},V^{>},\pp(v))\bigr )(\ell)\,&=\,\chi(\ell) \biggl (   
\Pi_\ell P_\ell F_2\bigl (\tilde v\bigr )_\ell
+ \Pi_\ell P_\ell F_3\bigl (\tilde v\bigr )_\ell\biggr )~,\cr
\bigl ( f^\s(V^{<},V^{>},\pp(v))\bigr )(\ell)\,&=\,\bigl
(1-\chi(\ell)\bigr ) \biggl (   
\Pi_\ell P_\ell F_2\bigl (\tilde v\bigr )_\ell
+ \Pi_\ell P_\ell F_3\bigl (\tilde v\bigr )_\ell\biggr )~,\cr
}
$$
and
$$
\tilde v _\ell(x)\,=\,(V^{<}(\ell)+V^{>}(\ell))\cdot \phi_\ell(x)+
\pp(v)_\ell(x)~.
$$
Note that since $V^{>}$ is supported outside $[-\ell_0,\ell_0]$, both
it and $\pp(v)$ decay exponentially (at least at the linear level)
and hence will be irrelevant for the asymptotics of $V^{<}$, as we shall
show. With this in mind, we introduce a new coordinate, $V^\s$, which
combines the ``irrelevant'' pieces, $V^\s=(V^{>},\pp(v))$. Then the
Eq.\equ(aggv) combined with Eq.\equ(ppv) takes the more suggestive
form
$$
\eqalign{
\partial_t V^{<}(\ell)\,&=\,\lambda_\ell  V^{<}(\ell) -
\bigl (f(V^{<},V^{\rm s})\bigr )(\ell)~,\cr
\partial_t V^{\rm s}\,&=\,\LL_b^{(0)} V^{\rm s} + g(V^{<},V^{\rm s})~,
}
\EQ(system)
$$
and we know that the spectrum of the linear operator $\LL_b^{(0)}$ is contained
in $(-\infty ,-\sigma^\s)$, for some $\sigma^\s>0$.

In order to proceed further, we analyze the non-linear terms in
Eq.\equ(system) in more detail. In particular, we concentrate on the
most critical terms, namely those in $f$ of Eq.\equ(system) which
depend only on $V^{<}$. We decompose $f(V^{<},V^\s)=f_2^{(0)}(V^{<})+f_3^{(0)}(V^{<})
+f_4^{(0)}(V^{<},V^\s)$,
where $f_2^{(0)}$ collects the terms which are homogeneous of
degree 2 in $V^{<}$
and $f_3^{(0)} $
those of degree 3. One gets
$$
\eqalign{
\bigl (f_2^{(0)}(V^{<})\bigr )(\ell)\,&=\,3 \chi(\ell)
\int dx\,\overline\phi _\ell(x)
u_\epsilon (x)\hide{e^{-i\ell x}}\int_{-1/2}^{1/2}\kern -1em dk\,\phi _k(x)\phi
_{\ell-k}(x) V^{<}(k)V^{<}(\ell-k)\cr
\,&\equiv\, 3 \chi(\ell)\int_{-1/2}^{1/2}\kern -1em dk\, K_2(\ell,k) V^{<}(k)
V^{<}(\ell-k)~,\cr
\bigl (f_3^{(0)}(V^{<})\bigr )(\ell)\,&=\, \chi(\ell)
\int dx\,\overline\phi _\ell(x)
\hide{e^{-i\ell x}}\int_{-1/2}^{1/2}\kern -1em dk_1\, dk_2\,\phi _{k_1}(x)\phi_{k_2}(x)
\phi_{\ell-k_1-k_2}(x)\cr
&~~~~~~~~~~~~\times  V^{<}(k_1)V^{<}(k_2)V^{<}(\ell-k_1-k_2)\cr
\,&\equiv\, \chi(\ell)\int_{-1/2}^{1/2}\kern -1em dk_1 \,dk_2\,
K_3(\ell,k_1,k_2)V^{<}(k_1)V^{<}(k_2)
V^{<}(\ell-k_1-k_2)~.\cr
}
\EQ(f1f2)
$$
At this point, we make use of the diffusive nature of the problem for
$V^{<}$, by introducing scaling variables
as in [W]. This will give us a more precise description of the
convergence process than the one obtained in [Sch]. We rescale the variables
in Eq.\equ(system) as follows: 
We first fix, once and for all, a (large) constant $t_0>0$. Then we define
$$
\eqalign{
V^{<}(\ell,t)\,&=\, w^\c \bigl ( \sign(\ell) \sqrt {|\Lambda_\ell |
\tttb\,}, \log \tttb \bigr )~,\cr
V^\s(\ell,t)\,&=\,w^\s \bigl (\sign(\ell) \sqrt {|\Lambda_\ell |
\tttb\,},\log \tttb\bigr )/\tttb^{1/2} ~,
}
\EQ(system2)
$$
where $\Lambda_\ell=\lambda _\ell$ for $|\ell|<\ell_0/2$ and is
monotonically extended beyond that region in such a way that it is
parabolic for large $|\ell|$. (This artifact is needed
because we have no guarantee that $\lambda _\ell$ itself is monotone.)
Note that if $\lambda_\ell$ were equal to $-\const\ell^2$,
this scaling would amount to the usual ``diffusive'' rescaling.
Our choice takes into
account higher order corrections produced by higher order terms in
$\lambda_\ell $. If we let now $p= \sign(\ell) \sqrt {|\Lambda_\ell |
\tttb}$, and $\tau =\log \tttb$, then Eq.\equ(system)
implies that $w^\c $ and $w^\s $ obey the following
equations:
$$
\eqalign{
\partial_\tau  w^\c \,&=\,(-p^2 -\HALF p\partial_p ) w^\c \cr
&~~ + e^{\tau}
\left (f_2(w^\c,e^{-\tau/2} )+f_3(w^\c,e^{-\tau/2})+f_4(w^\c ,w^\s e^{-\tau/2},e^{-\tau/2} )\right )~,\cr
e^{-\tau}\partial_\tau  w^\s\,&=\,  M_{\exp(-\tau/2)} w^\s + \HALF
e^{-\tau}w^\s
-\HALF e^{-\tau } p\partial_p w^\s+
e^{\tau/2}g(w^\c ,w^\s e^{-\tau/2},e^{-\tau/2} )~,\cr 
}
\EQ(wequ)
$$
where $f_2$, $f_3$, $f_4$ and $M$ in Eq.\equ(wequ) are defined below.
If
$$
pe^{-\tau /2}\,=\,p\tttb^{-1/2} \,=\,\sign(\ell )\sqrt{| \Lambda_\ell |}~,
$$
and if we denote
the inverse transformation by
$$
\ell\,=\,\Phi(pe^{-\tau/2 })~,
$$
where $\Phi$ is the inverse function of $x\mapsto
\sign(x)\sqrt{|\Lambda_x|}$,
then, given a function $w=w(\ell,t)$, we define the 
nonlinearity
$$
\eqalign{
\bigl [f_2(w,e^{-\tau/2})\bigr ](p)\,&=\,\bigl
[f_2^{(0)}(w(\cdot,e^\tau))\bigr ]
(\Phi(pe^{-\tau/2 }))\cr\,&=\,\bigl
[f_2^{(0)}(w(\cdot,\ttt))\bigr ]
(\Phi(p\tttb^{-1/2}))~.
}
$$
(Note that $\Phi(x)=x \bigl (1+\OO(x))$.) Analogous definitions apply
to $f_3$ and $f_4$. The operator $M$ will be described in detail in Eq.(2.13).
\REMARK The non-linearities $f_2$,\dots depend on the choice of
$t_0$. If we consider the initial value problem for the
Swift-Hohenberg equation, the ``smallness'' assumption on the
perturbation of the periodic state is to be understood with respect to
a choice of a (sufficiently large) $t_0$. As we will see, however,
the nonlinear terms can be bounded, independent of $t_0$, for
all $t_0 \ge T > 0$.

To this change of variables will correspond the following
(non-exhaustive)
list of substitutions in the integrals in Eq.\equ(f1f2):
Let $a,b\in[-\HALF,\HALF]$. Then
$$
\eqalign{
\chi(\ell)\int_a^b dk\, ~~&\rightarrow ~~ \chi\bigl(\Phi(p e^{-\tau /2}) 
\bigr )
e^{-\tau
/2}\int_{e^{\tau/2}\Phi^{-1}(a)}^{e^{\tau/2}\Phi^{-1}(b)}
 dq\,\Phi'(q e^{-\tau /2 })~,\cr
\phi_\ell ~~&\rightarrow ~~ \phi_{\Phi(p e^{-\tau /2})}~,\cr
\phi_{k-\ell} ~~&\rightarrow ~~
\phi_{\Gamma(p,q,\tau)} ~,\cr 
V(k,t)~~&\rightarrow ~~ w(p,\tau)~,\cr
V(\ell-k,t)~~&\rightarrow ~~ w(\Delta(p,q,\tau))~.\cr
}\EQ(x0)
$$
Here,
we define
$$
\eqalign{
\gpq\,&=\,\Phi(pe^{-\tau /2})-\Phi(qe^{-\tau/2})~,\cr
\dpq \,&=\,e^{\tau/2}\Phi^{-1}\bigl ( \Phi(pe^{-\tau
/2})-\Phi(qe^{-\tau/2})\bigr )~.
}
\EQ(x00)
$$
It follows at once from the definition of $\Phi$ that 
$$
\eqalign{
\gpq\,&=\,e^{-\tau/2}(p-q)\cdot\bigl (1+\gamma(p,q,\tau)\bigr )~,\cr
\dpq\,&=\,(p-q)\cdot\bigl (1+\kappa(p,q,\tau)\bigr )~,
}
\EQ(x000)
$$
where $\kappa$ and $\gamma$ are bounded and smooth.

We next discuss in detail the {\em spectrum} of $M_{\exp(-\tau/2)} $, which
is just the rescaled linear operator for the ``stable'' part of
$w$, {\it cf.}~Eq.\equ(system).
Recall first that $V^\s=(V^{>},\pp(v))$. This introduces a natural
decomposition of $w^\s=(w^\s_1,w^\s_2)$, as well as of
$M_{\exp(-\tau/2)}
=M_{{\exp(-\tau/2)},1}\oplus
M_{{\exp(-\tau/2)},2}$.
From the definition of the first component, we get
$$
\bigl (M_{\exp(-\tau /2) ,1} f^\s_1\bigr )(p,\tau )\,=\,\biggl(\epsilon^2 - \bigl
(1+(i+i\Phi(pe^{-\tau/2}))^2\bigr )^2-\KK\bigl (\Phi(pe^{-\tau/2})\bigr )\biggr ) f^\s_1 (p,\tau )~,
\EQ(m1def)
$$
where $\KK(\ell) $ is a kernel given by  
$$
\KK(\ell)\,=\,3\int dx\, \overline \phi_{\ell}(x) \hide{e^{-i\ell x}} u_\epsilon^2(x)
\phi_{\ell} (x)~.
$$ 
(Recall that $\phi_\ell$ really depends on $\epsilon $ as well and
should be written $\phi_{\epsilon ,\ell}$.)
Since $V^\s$ has support bounded away from $\ell=0$, say
$|\ell|>\ell_0/2$, we see that
$w^\s_1(p,\tau)$ will have support in $|p|e^{-\tau/2}>\sqrt
{|\Lambda_{\ell_0/3} |
}$, and the spectrum of $M_{\exp(-\tau/2),1}$ is seen to be contained in $\{\sigma |
\Re \sigma \le \sigma_0<0\}$, for some $\sigma_0$ and for all
$\tau>0$.

A very similar argument detailed in Appendix B shows that
the spectrum of $M_{{\exp(-\tau/2)},2}$ is also contained in such a set. Thus,
{\em the linear evolution generated by $M_{\exp(-\tau/2)} $ contracts
exponentially}. See Lemma B.6 below for details.

We next consider the operator $L=(-p^2 - \HALF p\partial_p )$, which
appears in the first component of Eq.\equ(wequ).
The detailed study of the semi-group 
generated by
$L$ will be given in Appendix B.
Here, we discuss its properties on an informal level.
The Fourier transform of $L$ is
$\partial_x^2+\HALF x\partial_x+\HALF$, which 
is conjugate to the harmonic oscillator
$H_0=\partial_x^2-x^2/16+1/4$
by the (unbounded!) transformation $T$, of multiplication
by $\exp(x^2/8)$. In formulas: $L=T^{-1} H_0 T$.
Therefore, $H_0$ has (say, on $L^2$), 
discrete spectrum 
$\mu_j=-j/2$, $j=0,1,\dots~$. 
It is this spectrum which leads to a nice interpretation of
the convergence
properties of the Swift-Hohenberg equation. The eigenvalues
of $L$ are unchanged by the transformation $T$, (and the
eigenfunctions are multiplied by a Gaussian), so
to each eigenvalue $\mu$ of $L$ there corresponds a decay
rate $e^{\tau\mu}$ in the linear problem.
Because of the transformation of variables
from $t$ to $\tau$, this decay rate becomes
$\tttb^{\mu}$ in the original problem Eq.\equ(system). 
In other words: Neglecting
the non-linearities in Eq.\equ(wequ) and setting $w^\s=0$, (and ignoring
potential problems related to the unbounded operator $T$)
we have a
solution
$$
w^\c (p,\tau) \,=\,\sum_{m=0}^\infty  w_m e^{-\tau m/2} \HH_m(2p)~,
\EQ(xx1)
$$
where $\HH_m$ is the $m^{\rm th}$ eigenfunction
of $L$. In the original
variables, this means that
$$
V^{<}(\ell,t)\,=\,\sum_{m=0}^\infty  w_m \tttb^{-m/2}
\HH_m\bigl (2\ell \tttb^{1/2} ( 1 + \OO(|\ell|^{1/2}) ) \bigr )~.
\EQ(adecay)
$$
Thus, {\em to each $m$ there corresponds a specific rate ($\mu_m=-m/2$) of decay} for
a part of
the function $V^{<}$.
Note that a change of $t_0$ just corresponds to a rearrangement of the
series.
(This is not contradictory, since a change of $t_0$ also changes the
initial condition, and hence the solution whose asymptotics
we are computing.)
In particular, the slowest rate of decay is associated with $\HH_0$,
which is Gaussian, and thus, at least at the linear level, a
``generic'' perturbation of the stationary state will decay like 
$\exp(-c\ell^2
t)$, for some $c>0$. In terms of the original independent variables
$(x,t)$, it decays like
$t^{-1/2}\exp(-x^2/(4tc))$, as $t\to\infty $.
This means that at this level, the periodic stationary states
are stable, and that perturbations of them decay like solutions
of the linear heat equation.
The invariant manifold theory guarantees that this behavior persists
in the non-linear problem, and in fact it tells us more. We will
see that in suitable spaces 
we can construct a sequence of manifolds $\MM_j$ of dimension
$j=1,2,\dots$, such that any solution of Eq.\equ(wequ) approaches a
solution on $\MM_j$ at a rate $e^{\tau \mu_{j-1}}$,
or again reverting to the original $(x,t)$ variables, at a rate
$\OO(\tttb^{\mu_{j-1}})$. In the case at hand, this is $\OO(\tttb^{-j/2})$.
Thus, in principle, we can analyze finer and finer details of the
asymptotics of perturbations of the stationary state by considering
the behavior of the solution on these {\em finite dimensional} manifolds.

\SECTION Casting the \clm(stab) into an invariant manifold theorem

At the end of the preceding section, we have seen that the spectrum of
the linear part of Eq.\equ(wequ) has the following nature: The
component $w^\c$ satisfies a differential equation
whose linear part has eigenvalues $\mu_j=-j/2$,
$j=0,1,\dots,N$, {\em provided} we work on a space of sufficiently
smooth and rapidly decaying 
functions. The evolution of $w^\s$ is governed by an equation
with an even more stable spectrum.

The invariant manifold theorem will show in which sense the built-in
scalings of Eq.\equ(xx1) 
survive the addition of non-linearities. While this presents
no conceptual problems at all---and this is the beauty of the present
approach---some care is of course needed in the application of the
invariant manifold theorem. Another point which might be overlooked is
the following: The invariant manifold theorem does {\em not} say that the
representation of 
the full non-linear problem is the same as in Eq.\equ(adecay),
but with slowly varying $w_j$. Rather, we will show 
that on the complement of a
dimension $j-1$ surface in the function space,
the solutions decay at least like $t^{-j/2}$, (for every $j\ge
1$), provided the initial data are sufficiently small and smooth.

In order to apply the invariant manifold method to the problem, we
need bounds on the non-linearities and bounds on the semi-group
generated by $L$. While the factor of $t=\exp(\tau)$ 
in front of $f_2$ in Eq.\equ(wequ) might look like
a disaster, we will see that by working in appropriate function spaces,
and taking advantage of the nature of the nonlinear term, this factor
will disappear. Its presence is in part due to the fact that
we chose to work in ``momentum'' space, rather than ``position''
space, because the linear problem is most naturally studied in Floquet variables.
If we rewrote these terms in position space ({\it i.e.}, in the original $(x,t)$
variables), they would look much less singular.

We will work in Sobolev spaces, and we define
$$
\Hqr\,=\, \{ v ~|~  (1-\partial_p^2)^{r/2}(1 + p^2)^{q/2} v \in L^2 \}~,
\EQ(hnorm)
$$
equipped with the corresponding norm $\|\cdot\|_{q,r}$.
The function $w^\c$ will be an element of $\Hqr$. 

The function
$w^\s$ has two components. The first component comes from
the central branch of the spectrum of the linear operator \equ(loperator),
and will also be in 
$\Hqr$. The second component comes from the stable
branches of the spectrum, and it depends on both $p$, and $x$.
It will be an element of the space:
$$
\eqalign{
\Hqrn \,=\, \{ w = w(p;x) &~ |~ w(p;x) = w(p;x+2 \pi), \cr 
& (1 - \partial_x^2)^{\nu/2} (1-\partial_p^2)^{r/2}(1+p^2)^{q/2} w 
\in L^2(\real \times [-\pi, \pi]) \} ~. 
\cr}
$$
By a slight abuse of notation, we will denote
by $\| w^\s \|_\HHqrn$ the sum of the $\Hqr $ norm of the 
first component of $w^\s$ and the $\Hqrn $ norm of the second component,
and by $\| w^\s\|_{q,r,\nu}$, we will mean the $\Hqrn $ norm of
just the second component. We will also use $\HHqrn$ to denote the 
space of all functions with finite $\HHqrn$ norm.

The non-linearities satisfy the following bounds:
\CLAIM Proposition(nonlin) For every $q\ge 2$ and every $r\ge 0$
there is a constant $C$ for which
$$
\eqalign{
\| e^{\tau} f_2( w,e^{-\tau/2}) \|_ {q-1,r}\,&\le\, C  \| w\| _{q,r}^2~,\cr
\| e^{\tau} f_3( w,e^{-\tau/2}) \|_ {q,r}\,&\le\, C \| w\|
_{q,r}^3~,\cr}
\EQ(fbound2)
$$
for all $\tau>0$.

\CLAIM Proposition(nonlin2) For every $q\ge 2$ and every $r\ge 0$
there is a constant $C$ for which
$$
\eqalignno{
\|e^{\tau}  f_4( w^\c,w^\s e^{-\tau /2},e^{-\tau/2}) \|_ {q,r}~&\cr
\,\le\,
Ce^{\tau /2}
\| w^\s\|_{\HHqrn }\,&\bigl (e^{-\tau /2}\|w^\c\|_{{q,r}}+e^{-\tau
}\|w^\s\|_{\HHqrn }\bigr )\cr
\times &\bigl (1+e^{-\tau /2}\|w^\c\|_{{q,r}}+e^{-\tau }\|w^\s\|_{\HHqrn }\bigr 
)~,\NR(fboundf4)
\|e^{\tau /2} g(w^\c,\,w^\s e^{-\tau /2},e^{-\tau/2 })\|_\HHqrn~&
\cr\,\le\,
Ce^\tau \bigl( e^{-\tau /2}\| w^\c\|_{q,r}&+e^{-\tau }\|
w^\s\|_{\HHqrn }\bigr )^2\cr
\times \bigl (1+ e^{-\tau /2}\|& w^\c\|_{q,r}+e^{-\tau }\|
w^\s\|_{\HHqrn }\bigr ) ~,\NR(fboundg)
}
$$
for all $\tau>0$.

\REMARK Note that every factor of $\|w^\c\|_{{q,r}}$ is multiplied by
$e^{-\tau /2}$ and every factor of $\|w^\s\|_\HHqrn$ is multiplied by
$e^{-\tau }$.

\REMARK As we pointed out above, the nonlinear terms depend on the
constant $t_0$.  However, the bounds in the two preceding
propositions are independent of $t_0$.  More precisely, for
any $T > 0$, the
the constants $C$ in both propositions can be chosen so that
the estimates in \equ(fbound2)--\equ(fboundg) hold for all $t_0 \ge T$.

The proofs will be given in Appendix A.
Note that one loses a power of $p$ in the first
estimate of Eq.\equ(fbound2), but of course,
one ``gains'' the square of the function.

We will regain the ``lost'' power of $p$ by examining in detail the
semi-group generated by $L$. 
We denote by $P_N$ the projection onto the space spanned by the
$N$ eigenvalues $\{\mu_j=-j/2\}_{j=0,\dots,N-1}$ of $L$.
We define $Q_N=1-P_N$. (We verify in Appendix B that these projections
are defined.)
On the space corresponding to $Q_N$, we expect 
the norm of the semi-group generated by $L$
to decay like $\exp(\tau \mu_N)$. This is indeed the case. 
\CLAIM Theorem(semigroup) For every $\epsilon >0$, 
there is a constant $N_0$ and a function $r(N,q)$
such that for every
$N\ge N_0 $, every $q\ge1$ 
and every $r\ge r(N,q)$, there
is a $C=C(q,r,N)<\infty $ such that
$$
\left \| e^{\tau  L} Q_N v\right \|_{q,r}\,\le\,
{C(q,r,N)\over \sqrt{a(\tau )}} e^{- \tau (|\mu_N|-\epsilon ) } \|v\|_{q-1,r}~,
\EQ(decay0)
$$
where $a(\tau )=1-e^{-\tau }$ and $L=-p^2-\HALF p \partial_p$

The proof will be given in Appendix B.

We also need an estimate on the linear evolution generated by
$M_{\exp(-\tau/2)}$. Let $U_{\tau}$ be the solution of
$$
e^{-\tau} \partial_\tau U_{\tau} \,=\, M_{\exp(-\tau/2)} U_{\tau}~,
$$
with initial condition $U_{0} = 1$.
(Compare with the linear part of \equ(wequ).) Then, in Appendix B, we
prove
\CLAIM Theorem(semigroup2) If $w_0 \in \HHqrn $, then there exists $c_0 >0$,
such that for all $\tau \ge 0$,
$$
\|U_{\tau} w_0\|_{\HHqrn} \,\le\, \exp(-e^{c_0 \tau/2}) \| w_0 \|_{\HHqrn}~.
$$

With the help of the bounds \clm(nonlin)--\clm(semigroup2),
we can now reformulate the problem in terms of invariant manifolds.
The Eq.\equ(wequ) can be written as an {\em autonomous system}
by defining $\eta = \tttb^{-1/2}=e^{-\tau/2}$:
$$
\eqalign{
\partial_\tau  w^\c \,&=\,L w^\c  + \eta^{-2}
\left (f_2(w^\c,\eta )+f_3(w^\c,\eta )+f_4(w^\c ,w^\s\eta,\eta) \right)~,\cr
\eta^2\partial_\tau w^\s\,&=\,
  M_\eta  w^\s+
\eta^{-1}g(w^\c ,w^\s\eta ,\eta)~,\cr 
\partial_\tau \eta\,&=\,-\HALF\eta~.\cr
}
\EQ(wequ2)
$$

We will construct an invariant manifold tangent at the origin to the
eigenspace corresponding to the $N$ largest eigenvalues of $L$, and the
$\eta$ direction.
We subdivide the center variable $w^\c$ according to the projection 
$Q_N$ defined earlier, where $N$ is fixed once and for all. 
Define
$$
x_1\,=\,(1-Q_N) w^\c,\quad x_2\,=\,Q_N w^\c, \quad x_3\,=\,w^\s~.
\EQ(decomp)
$$
Note that the variable $x_1$ is in a finite dimensional space, while $x_2$
and $x_3$ are in infinite dimensional Hilbert spaces.
The system of equations Eq.\equ(wequ2) now takes the form
$$
\eqalign{
\partial_\tau x_1 \,&=\, A_{1} x_1 + N_1( x_1,\eta, x_2 ,x_3)~,\cr
\partial_\tau \eta\,&=\, -\HALF\eta~,\cr
\partial_\tau x_2 \,&=\, A_{2} x_2 + N_2( x_1,\eta, x_2 ,x_3)~,\cr
\eta^2\partial_\tau x_3 \,&=\, A_{3,\eta} x_3 + N_3( x_1,\eta, x_2 ,x_3)~.\cr
}
\EQ(full)
$$
Here $A_1=(1-Q_N) L$, $A_2=Q_N L$, and $A_{3,\eta}=M_\eta$.
\REMARK In view of later developments, we consider $x_1$ and $\eta$ to
be the ``interesting'' variables and $x_2$ and $x_3$ the ``slaved''
variables, hence the new order of the variables.

\REMARK Eq.\equ(full) is a very singular perturbation problem, because
of the factor 
of $\eta^2$ in front of the derivative of $x_3$. What is more, since $\eta(\tau)
= e^{-\tau/2}$, it becomes steadily more singular in precisely the limiting
regime in which we are interested. Nonetheless, we will see that the invariant
manifold theorem provides just the tool we need to understand this limit. 
Singular perturbation problems of this type do not seem to have been studied
much, but they do arise naturally in other contexts, such as the study of 
parabolic equations in cylindrical domains ([W2]).

We shall call Eq.\equ(full) the {\em full system}. 
To simplify the notation,
we shall omit the dependence on $\eta $ in $A_{3,\eta }$. Consider the
spectra of $A_1$, $A_2$, $A_3$. From what we have seen earlier, we
find that 
$$
\eqalign{
{\rm spec} (A_1)  \,&=\,  \{ 0, -1/2, -1, \dots ,-(N-1)/2 \} ~,\cr
{\rm spec} (A_2)  \,&\subseteq\,[-\infty ,-N/2]~,\cr
{\rm spec} (\eta ^{-2}A_3)  \,&=\, [-\infty ,-c/\eta^2 ] ~,\cr
}
\EQ(spectrum)
$$ 
where $c$ is some positive constant. Thus, we expect to apply a pseudo
center manifold theorem to ``slave'' the variables $x_2$, $x_3$ to the
variables $x_1$ and $\eta$. While there are certain technical difficulties
associated with the very singular perturbation, in Appendix C, 
we demonstrate 
the following Proposition:
\CLAIM Proposition(invariantman) Fix $N > 0$. There exist $r > 0$, $q \ge 1$,
and $\nu > 1/2$, such that the system of equations \equ(full) has an
invariant, $N+1$-dimensional manifold, given in a neighborhood of the
origin by the graph of a pair of functions 
$$
\eqalign{
h^*_2: \real^N \times \real & \to \Hqr ~,\cr h^*_3: \real^N \times
\real & \to \HHqrn~.\cr}
$$

We next turn to the task of showing that the invariant manifold we
found for Eq.\equ(wequ2) actually attracts solutions at an
exponential rate.
\LIKEREMARK{Notation}It is useful to introduce the notation
$\xi=(x_1,\eta)$ for the two
relevant variables.

Consider a solution of the form $\bigl(w^\c(\tau ),w^\s(\tau )\bigr)$
of Eq.\equ(wequ2), with $w^\c(\tau )=\bigl (x_1(\tau ),x_2(\tau
)\bigr )$ as in
Eq.\equ(decomp), and $w^\s(\tau )=x_3(\tau )$.
We wish to show that 
$$
\bigl (\,\xi(\tau )\,,\,x_2(\tau )\,,\,x_3(\tau )\,\bigr )  ~~\longrightarrow
\bigl (\,\xi(\tau )\,,\,h_2^*(\xi(\tau ))\,,\,h_3^*(\xi(\tau ))\,
\bigr )~,
$$
as $\tau \to\infty $, and furthermore, that it does so at an
exponential rate, given essentially by the least negative eigenvalue,
$\mu_N$,
of the operator $A_2$.
\CLAIM Proposition(522962) Fix $N>0$. For every $\delta $ satisfying 
$0<\delta$ there is an $\epsilon _0>0$ 
such that if the solution of Eq.\equ(wequ2)
remains in a neighborhood of the
origin of size $\epsilon _0$ one has the following bound:
There is a $C^*<\infty $ for which
$$
\bigl \|x_2(\tau )-h_2^*(\xi(\tau ))\|_{q,r} +\|x_3(\tau
)-h_3^*(\xi(\tau ))\|_{\HHqrn}\,\le\,
C^* e^{-(|\mu_N|-\delta)\tau } ~,
$$
as $\tau \to\infty $.

\PROOF This proof is relatively standard, see {\it e.g.}, Carr [C]. 
Let 
$$
z(\tau )\,=\,\left ({x_2(\tau )-h_2^*(\xi(\tau ))\atop x_3(\tau
)-h_3^*(
\xi(\tau )\bigr )}\right)\,\equiv\,
\left ({z_2(\tau )\atop z_3(\tau )}\right )~.
$$
Then we have
$$
\dot z \,=\,\left({A_2 z_2 +\hat N_2(\xi,z_2,z_3) \atop \eta^{-2} A_3 z +\eta^{-2}
\hat N_3(\xi,z_2,z_3 )}\right )~,
\EQ(z1)
$$
where, 
with the notation of Eq.\equ(full), 
$$
\hat N_j(\xi,z_2,z_3 )\,=\,N_j(\xi,z_2+h_2^*(\xi ),z_3
+h_3^*(\xi))
\,\,-\,\,N_j(\xi ,h_2^*(\xi ),h_3^*(\xi ) )~,
$$
for $j=2,3$.
The only novelty in Eq.\equ(z1) 
w.r.t.~[C] is the factor of $\eta^{-2}$ in
the ``3''-component which is the reason for our repeating his arguments.
But we can integrate Eq.\equ(z1) explicitly and get
$$
\eqalign{
z_2(\tau )\,&=\,e^{\tau A_2} z_2(0)+\int_0^\tau  d\sigma\,
e^{(\tau -\sigma)A_2} \hat N_2\bigl (\xi(\sigma),z_2(\sigma),z_3(\sigma)\bigr
)~,\cr
z_3(\tau )\,&=\,e^{(\eta(\tau )^{-2}-\eta (0)^{-2})A_3 }z_3(0)
+\int_0^\tau d\sigma\,
{1\over \eta(\sigma)^2} e^{(\eta(\tau )^{-2}-\eta (\sigma)^{-2})A_3} \hat N_3\bigl (\xi(\sigma),z_2(\sigma),z_3(\sigma)\bigr
)~.\cr
}
$$
We assume $\eta(0)>0$, since we are interested in the case
$\eta(0)=t_0^{-1/2}$, and we have chosen the scaling factor $t_0$ to
be a positive, finite constant.
Note also that $\xi$ remains in a neighborhood of the origin, as $\tau
\to\infty$.
From the bounds on the non-linear terms we see that if the solution
satisfies
$$
\|x_2(\tau )\|_{{q,r}}+\|x_3(\tau )\|_{\HHqrn }\,\le\, \rho ~,
$$
for all $\tau \ge0$, then, with $\nu_N=N/2$, the modulus of the $N^{\rm th}$
eigenvalue $\mu_N$ of $L$, we have
$$
\eqalign{
\|z_2(\tau )\|_{q,r}\,&\le\,e^{-\tau \nu_N}\|z_2(0)\|_{q,r}
+C\epsilon \int_0^\tau  d\sigma e^{-(\tau -\sigma)\nu_N} \bigl (
\|z_2(\sigma)\|_{q,r} +\|z_3(\sigma)\|_{\HHqrn }\bigr )~,\cr
\|z_3(\tau )\|_{\HHqrn }\,&\le\,e^{(\eta(\tau )^{-2}-\eta (0)^{-2})\nu_N}\|z_3(0)\|_{\HHqrn }\cr
&~~~+C\epsilon
 \int_0^\tau  d\sigma\, {1\over \eta(\sigma)^{2}} e^{-(\eta(\tau )^{-1}-\eta (\sigma)^{-1})\nu_N} \bigl (
\|z_2(\sigma)\|_{q,r} +\|z_3(\sigma)\|_{\HHqrn }\bigr )~.\cr
}
\EQ(z5)$$
In deriving these inequalities, we used the inequalities
$$
\eqalign{
\|e^{\tau A_2} \hat N_2(\xi,z_2,z_3 )\|_{q,r}
\,&\le\,e^{-\tau \nu_N}\|\hat N_2(\xi,z_2,z_3)\|_{q-1,r}~,\cr
\|e^{\rho A_3  } \hat N_3(\xi,z_2,z_3 )\|_{\HHqrn }\,&\le\,e^{-\rho
\nu_N}\| \hat N_3(\xi,z_2,z_3)\|_{\HHqrn }~,\cr}
$$
which follow
from the bounds of Appendix B.
If we now fix $\delta >0$ and define
$$
\eqalign{
C_2(\tau) \,&=\,\sup_{0\le \tau '\le \tau } e^{\tau' (\nu_N-\delta)}
\|z_2(\tau ')\|_{q,r}~,\cr
C_3(\tau) \,&=\,\sup_{0\le \tau '\le \tau } e^{\tau' (\nu_N-\delta)}
\|z_3(\tau ')\|_{\HHqrn }~,\cr
}
$$
then the Eq.\equ(z5) leads to the inequality
$$
\eqalign{
C_2(\tau )\,&\le\,K_1 + K_2 \epsilon  \bigl (C_2(\tau )+C_3(\tau
)\bigr )
\,\int_0^\tau d\sigma\, e^{-(\tau-\sigma)\delta}~,\cr  
C_3(\tau )\,&\le\,K_3 + K_4 \epsilon  \bigl (C_2(\tau )+C_3(\tau
)\bigr )
\,\int_0^\tau d\sigma\, {1\over \eta(\sigma)^2} e^{(\eta(\tau
)^{-2}-\eta(  \sigma)^{-2})\nu_N}e^{(\tau-\sigma)(\nu_N-\delta)}~.\cr  
}
$$
If we insert into these integrals the definitions 
$$\eta(\sigma)
\,=\, \exp(-\sigma/2) \eta(0)~,\quad\eta(\tau) \,=\,\exp(-\tau/2)
\eta(0)~,
$$ 
we 
find that 
both integrals are uniformly bounded in $\tau \ge0$ if $\eta(0) $ is
in a compact subinterval of $(0,1)$. The proof of \clm(522962) is complete.

Thus, all solutions near the invariant manifold approach it
exponentially fast in $\tau $.

One can now show without difficulty that every solution approaches
exponentially quickly a {\em particular} solution on the (approximate)
invariant manifold 
$$\bigl (x_1(\tau ),\eta=0,h_2^*(x_1(\tau
),0),h_3^*(x_1(\tau),0)\bigr )~.$$
This
consists simply in translating the pp.21--24 of [C] into the
present setting and thus there is no need to repeat this argument here.

If we combine these results with \clm(invariantman), we arrive finally
at a description of the invariant manifolds which exist close to the
origin for \equ(full).
\CLAIM Theorem(full) Fix $N > 0$ and $\delta>0$. 
There exist $r > 0$, $q \ge 1$,
and $\nu > 1/2$, such that the system of equations \equ(full) has an
invariant, $N+1$-dimensional manifold, given in a neighborhood of the
origin by the graph of a pair of functions $h^*_2:
\real^N \times \real \to \Hqr$, and $h^*_3:
\real^N \times \real \to \HHqrn $. Any solution of \equ(full) which remains
in a neighborhood of the origin for all $\tau \ge 0$ approaches a solution of
the $N+1$-dimensional system of ordinary differential equations
$$
\eqalign{
\partial_\tau x_1 \,&=\, A_{1} x_1 +
	N_1( x_1,\eta, h^*_2(x_1,\eta) ,h^*_3(x_1,\eta))~,\cr
\partial_\tau \eta\,&=\, -\HALF\eta~,\cr
}
\EQ(restricted)
$$
which results from restricting \equ(full) to this invariant manifold.
Furthermore, the rate of approach to this manifold is
${\OO}(\exp(-\tau(N/2 - \delta)))$.

\REMARK This theorem almost suffices to prove \clm(stab). In particular,
it emphasizes that in a neighborhood of the periodic solutions of \equ(sh)
there exists a family of invariant manifolds, $M_2$, $M_3$, $\dots$, described
in that theorem. The one remaining piece of the puzzle is to describe
the behavior of solutions restricted to the invariant manifold, and
that we do in the next section.

\SECTION The projection of the non-linearity onto zero momentum

We have already shown that there exists a (smooth) invariant manifold,
parameterized by $(\xi,h_2^*(\xi),h_3^*(\xi))$, where $\xi=(x_1,\eta
)$. This manifold satisfies the equation Eq.\equ(full), which, in the
case of $N=1$, {\it i.e.}, in the case of a two-dimensional invariant
manifold amounts to
$$
\eqalign{
\partial_\tau x_1\,&=\,N_1\bigl (x_1,\eta ,h_2^*(\xi),h_3^*(\xi)\bigr
)~,\cr
\partial_\tau\eta \,&=\,-\HALF \eta ~,\cr
\partial_\tau \bigl (h_2^*(\xi)\bigr )\,&=\,
A_2 h_2^* (\xi) + N_2\bigl (x_1,\eta ,h_2^*(\xi),h_3^*(\xi)\bigr
)~,\cr
\eta^2 \partial_\tau \bigl (h_3^*(\xi))\,&=\,A_3 h_3^* (\xi)+N_3\bigl (x_1,\eta ,h_2^*(\xi),h_3^*(\xi)\bigr
)~.\cr}
\EQ(IM)
$$
Note that because $N=1$ the operator $A_1$ equals zero (which is the
highest eigenvalue of $L$).

To understand the dynamics inside this invariant manifold, we now
state and prove the following proposition, which is based on
Schneider's beautiful observation:
Let $\tilde N_1(x_1,\eta )$ be the r.h.s.~of the first equation in
\equ(IM), {\it i.e.}, $\partial_\tau x_1=\tilde N_1(x_1,\eta )$.
\CLAIM Proposition(project) There is an $x_{1,0}>0$ such that
$\tilde N_1(x_1,0)=0$, for all
$|x_1|<x_{1,0}$.

Thus, the non-linearity {\em vanishes identically} at ``infinite
time,'' which corresponds to $\eta=0$.
Before proving \clm(project), we show that it implies the following
important 
\CLAIM Theorem(fixed) If $x_1(0)$ is sufficiently close to 0, then
there are a constant $C<\infty $ and an $x_1^*$ such that 
$$
|x_1(\tau )-x_1^*|\,<\, Ce^{-\tau /2}~.
\EQ(converg)
$$

\PROOF Using the fact that $\eta(\tau)=e^{-\tau/2}$, we can rewrite
the equation for $x_1$ as
$$
\partial_\tau x_1 \,=\, \tilde{N}_1 (x_1, e^{-\tau/2})~.
\EQ(newN-eqn)
$$
Since $\tilde{N}_1$ is a smooth (at least $\CC^{1+\alpha}$) function
with $\tilde{N}_1(x_1,0) = 0$ in some neighborhood of the origin, there
exists a constant $C_N > 0$, such that 
$|\tilde{N}_1 (x_1, e^{-\tau/2})| \le C_N \exp(-\tau/2)$, for
$|x_1|$ sufficiently small. Integrating \equ(newN-eqn)
and applying this estimate yields:
$$
\eqalign{
|x_1(\tau_{\rm f}) - x_1(\tau_{\rm i})| \,&=\,
\left|\int_{\tau_{\rm i}}^{\tau_{\rm f}} d\sigma\,\tilde{N}_1(x(\sigma),e^{-\sigma/2}) \right|\cr
\,&\le\, C_N \int_{\tau_{\rm i}}^{\tau_{\rm f}} d\sigma \,e^{-\sigma/2} 
\,=\,2 C_N e^{-\tau_{\rm i}/2}(1-e^{(\tau_{\rm i}-\tau_{\rm f})/2})~.
\cr
}$$
This estimate immediately implies the behavior claimed in \clm(fixed).
\LIKEREMARK{Proof of \clm(project)}The basic idea is to relate $\tilde
N_1(x_1,0)$ to the non-linear term of {\em another problem}, which is
known to be 0. This other problem is the center manifold equation for
the perturbations of a stationary solution of Eq.\equ(sh) {\em
restricted to a space of $2\pi$-periodic functions}. In this case, the
equation analogous to Eq.\equ(v1)
is 
$$
\partial_t v\,=\, L_{\rm per} v + F(v)~,
$$
where $F(v)$ collects the non-linear terms in $v$. The spectrum of
$L_{\rm per}$ is pure point, with a simple zero eigenvalue, and all
others negative, and bounded away from 0. The eigenvector with 0
eigenvalue is $u_\epsilon '$, where $u_\epsilon $ is given by
Eq.\equ(sepsilon). 
If we call $x_{1,{\rm per}}$ the coordinate in the $u_\epsilon'$
direction, then there exists a one-dimensional center manifold,
tangent to this direction and given as the graph of a function
$H(x_{1,{\rm per}})$. A very nice observation by Schneider is that this
center manifold must coincide with the translates of the stationary
state $u_\epsilon $, which is formed of fixed points of the
Swift-Hohenberg Eq.\equ(sh). Hence, on this center manifold we must
have
$\dot x_{1,{\rm per}}=0$.
Using this information, the equations for this center manifold take a
particularly simple form.
Let $P_{{\rm per}}$ denote the projection onto $u'_\epsilon $ and
let $Q_{{\rm per}}=  1-P_{{\rm per}}$. Then the preceding discussion
implies that the flow $\psi_{t,{\rm per}}$ is the identity on
$x_{1,{\rm per}}$, and hence the equations
for the invariant manifold read:
$$
\eqalignno{
\dot x_{1,{\rm per}}\,&=\, P_{{\rm per}}F(x_{1,{\rm per}},H(x_{1,{\rm
per}}))\,=\,0~, 
\NR(per1)
H(x_{1,{\rm per}})\,&=\,\int_{-\infty }^0 d\tau \,e^{-
Q_{{\rm per}}L_{{\rm per}}} Q_{{\rm per}}F(x_{1,{\rm per}},H(x_{1,{\rm
per}})) \cr
\,&=\,-  \bigl (Q_{{\rm per}}L_{{\rm per}}\bigr )^{-1}
Q_{{\rm per}}F(x_{1,{\rm
per}},H(x_{1,{\rm 
per}}))~. \NR(per2)
}
$$
We now wish to use this information to prove \clm(project). The
rough idea is to show that
$$
\tilde N_1(x_1,0)\,=\,P_{\rm per} F(x_{1,{\rm per}},H(x_{1,{\rm
per}}))~,
\EQ(toshow)
$$
and this quantity vanishes by Eq.\equ(per1).
More precisely, we shall show:
\CLAIM Proposition(x3) The cubic term in $x_1$ of $\tilde
N_1(x_1,\eta)$ coincides in the limit $\eta\to0$ with the cubic term
in $x_1$ of $P_{\rm per} F(x_{1,\rm per}, H(x_{1,\rm per}))$. All other
terms in $\tilde N_1$ go to 0 as $\eta\to0$. 

\REMARK Since $P_{\rm per} F\bigl(x_{1,\rm per}, H(x_{1,\rm per})\bigr
)=0$, this
proves Eq.\equ(toshow) and thus \clm(project).
\PROOF The proof of \clm(x3) will be given in Appendix D.
\SECTION Completion of the proof of \clm(stab)

We now consider exactly how the results of the previous two
sections about the behavior of solutions in, and near, the 
invariant manifold translate back into statements about solutions
in terms of the original variables. We will focus specifically
on the case considered in the previous section in which the
invariant manifold is two-dimensional, with coordinates $(x_1,\eta)$,
but the results can be immediately extended to the case
of a manifold of arbitrary dimension.

Suppose we have a solution 
$w_\tau  = w^\c_\tau  + w^\s_\tau $,
of the system \equ(wequ2),
which remains in a neighborhood of the origin for all $\tau \ge 0$.
This will be the case if its initial condition is sufficiently
small in $H_{q,r} \oplus \HHqrn$.
We measure the size of $w$ in the norm $\I  \cdot \I $,
which is the sum of the $H_{q,r}$ norm of $w^\c$, and
the $\HHqrn$ norm of $w^\s$. 
By the results of \clm(full), we know that there exists
a solution, $w^{\rm inv}_\tau $, on the invariant manifold
such that 
$$
\I  w_\tau  -  w^{\rm inv}_\tau\I  \,\le\,
C e^{-\tau(1/2-\delta) }~,
\EQ(new1)
$$
with $\delta >0$.
In addition, from \clm(fixed), we know that there exists
some $w^*$, which lies in the invariant manifold
for which 
$$
\I  w^{\rm inv}_\tau  - w^* \I  \,\le\, C e^{-\tau/2}~.
\EQ(new2)
$$
Here, $w^*$ is the function whose coordinates in the invariant
manifold representation is just the limiting point
$x_1^*$ in \clm(fixed), {\it i.e.}, $w^*=\bigl
(x_1^*,0,h^*_2(x_1^*,0),h^*_3(x_1^*,0)\bigr )$. 
Combining \equ(new1) and \equ(new2), we see that for solutions
that remain near the origin, there exists a function $w^*$,
for which
$$
\I  w_\tau  -w^* \I  \,\le\, C e^{-\tau(1/2-\delta )}~.
\EQ(new3)
$$

Our final task is now to untangle the various changes of variables
which we made in the original equation.
If we first ``undo'' the rescaling in \equ(system2), we see that the solution
$v(\ell,t)$, corresponding to $w(\cdot,\tau)=w_\tau $ is
$$\eqalign{
v(\ell,t) \,&=\,w^\c(\sign(\ell)\sqrt{|\Lambda_{\ell}| \tttb},\log
\tttb)\cr
\,&+\, 
{{1}\over{\tttb^{1/2}}} w^\s( \sign(\ell)\sqrt{|\Lambda_{\ell}| \tttb},\log \tttb) \cr
\,& \equiv\, v^\c(\ell,t) + v^\s(\ell,t)~.\cr}
\EQ(new4)
$$
One can make a corresponding decomposition of $v^*$, the solution
corresponding to $w^*$.

First consider $v^\c$. From \equ(new3), one has
$$
\| w^\c_\tau  - w^{*,\c}_\tau \|_{q,r}^2 \,=\,
\int dp\, |(1 - \partial_{p}^2)^{r/2} (1+p^2)^{q/2} (w^\c(p,\tau ) -
w^{*,\c} (p,\tau ) )|^2 
\,\le\, C e^{-\tau(1-2\delta )}~.
\EQ(new5)
$$
According to \equ(new4), 
$w^\c(\ell,\tau) = v^\c(\Phi^{-1}(p e^{-\tau/2}),t)$, so
substituting this expression---and the analog for $w^{*,\c}$---into \equ(new5)
one finds that the left hand side of that inequality is equal to:
$$\eqalign{
\int dp\,&|(1 - \partial_{p}^2)^{r/2}   (1+p^2)^{q/2} 
\bigl (v^\c(\Phi^{-1}(p e^{-\tau/2}),t)
- v^{*,\c}(\Phi^{-1}(p e^{-\tau/2}),t)\bigr )  |^2  \cr
\,&\ge\, \int dp\,| (1+p^2)^{q/2} \bigl (v^\c(\Phi^{-1}(p e^{-\tau/2}),t)
- v^{*,\c}(\Phi^{-1}(p e^{-\tau/2}),t) \bigr ) |^2  \cr
\,&\ge\, \int d\ell\, \tttb^{1/2} \Phi'(\ell ) \,|(1+ \tttb(\Phi(\ell ))^2)^{q/2} \bigl(v^\c(\ell ,t) - v^{*,\c}(\ell ,t)\bigr)|^2  ~,\cr}
\EQ(new6)
$$
where in the last integral we changed the integration variable to 
$\ell = \Phi^{-1}(p e^{-\tau/2})= \Phi^{-1}(p \tttb^{-1/2})$.
\REMARK We dropped the derivatives with respect to $p$ in the second line of
\equ(new6) for simplicity---one could retain them at the expense of complicating
the following expressions.

Since $\Phi(x) \approx x$, for $x$ small, and is equal to a constant times $x$
for $|x|$ large (due to the definition of $\Lambda_{\ell}$), we see that combining
\equ(new5) and \equ(new6) and recalling that $t_0>0$, one finds:
$$
\int d\ell\,|(1+ \ell^2)^{q/2} (v^\c(\ell,t) - v^{*,\c}(\ell,t)) |^2
\,\le\, C t^{-3/2(1-2\delta )}~.
\EQ(new7)
$$

Analogous estimates hold for the ``stable'' part of the solution. Proceeding as above, 
one can show that 
$$
\sum_n (1+n^2)^\nu \int d\ell\,| (1+\ell^2)^{q/2} (v^\s(\ell,t)  -
v^{*,\s}(\ell,t) )|^2 \,\le\, C t^{-5/2(1-2\delta )}~.
\EQ(new8)
$$
Thus, the ``stable'' part of a solution near the origin approaches the solution
$v^*$ on the invariant manifold faster than the ``center'' part of the
solution. (An effect that is entirely in accord with one's intuition.)

We next take a closer look at the solution $w^*$ (or $v^*$) on the invariant
manifold. From the computation in the previous section, we know
that since the eigenfunction in the $x_1$ direction is $\exp(-p^2)$,
{\it cf.}~Eq\equ(xx1), we have
$w^*(p) = c^* \exp(-p^2) + h_3^*(c^*\exp(-p^2))$. If we now rewrite
this in terms of the $v(\ell,t)$ variables, we find
$$
v^*(\ell,t) \,=\, c^* e^{-\Lambda_\ell t} + t^{-1/2} h_3^*( c^* e^{-\Lambda_\ell t} )~.
\EQ(new9)
$$
Thus, if $v(\ell,t)$ is a solution of \equ(v1) (in the unscaled variables), we see from
\equ(new7)--\equ(new9) that 
in the $L^2( (1+\ell^2)^{q/2} d\ell)$ norm,
$$
v(\ell,t) \,=\, c^* e^{-\Lambda_\ell t} + \OO (t^{-1/2(1-2\delta )} )~.
\EQ(new10)
$$
But we know from Section 2 that $\Lambda_\ell  = \ell ^2 + \OO(\ell ^3)$ for $\ell $ small,
and $\Lambda_\ell  = c \ell ^2$, for $|\ell |$ large, so one finds by an easy and explicit
estimate that
$$
\int d\ell\,|(1+\ell^2)^{q/2}( e^{-\Lambda_\ell t} -  e^{-\ell ^2 t}) |^2 \,\le\, C t^{-1/2} ~.
\EQ(new11)
$$
Combining \equ(new10) and \equ(new11) one has
\CLAIM Proposition(asymptotics) If $v$ is a solution of \equ(v1) with sufficiently
small initial condition (in $H_{q,r}\oplus \Hqrn$), then
$$
(\int d\ell\, |(1+\ell ^2)^{q/2}( v(\ell ,t) - c^*e^{-\ell ^2 t})|^2
)^{1/2} \,\le\, C t^{-1/4(1-2\delta )} ~.
$$

Note that if we transform back to the 
$(x,t)$ variables, this implies the asymptotic
estimate in \clm(stab), and hence the proof of that theorem is complete.
\def\actualnumber{A}
\SECTION Bounds on the non-linearities

In this section, we prove \clm(nonlin) and \clm(nonlin2).
We begin by studying the kernels $K_2(\ell,k)$, and $K_3(\ell,k)$
introduced in \equ(f1f2).
\CLAIM Lemma(k2k3) There is a constant $C$ such that 
$$
|K_2(\ell,k)|\,\le\, C\epsilon \max\bigl( (|k|^2+|\ell|^2), 1\bigr)~.
$$

\PROOF By the definition of Eq.\equ(f1f2), we have
$$
K_2(\ell,k)\,=\,\int dx\,\hide{e^{-i\ell x}}\overline \phi_\ell(x) u_\epsilon
(x)\phi_k(x)\phi_{\ell-k}(x)~.
\EQ(k222)
$$ 
Since $u_{\epsilon}$ and $\phi_k$ are both uniformly bounded, we have immediately that
$|K_2(k,\ell)| \le C \epsilon$.
The crucial observation of Schneider[Sch] is that because of
Eq.\equ(phiell), repeated here for convenience 
$$
\phi _{\epsilon,\ell}(x) \,=\,u_\epsilon '(x)+i\ell g_\epsilon (x) +
h_{\epsilon,\ell } (x) \ell^2~,
\EQ(phiell2)
$$
(with real $g_\epsilon $), $K_2$ has an expansion 
$$
\int dx\, \hide{e^{-i\ell x}} u_\epsilon (x) (u_\epsilon'(x))^3
\,\,+\,\,u_\epsilon (x)\bigl (u_\epsilon'(x)\bigr )^2 \bigl (-i\ell +ik
+i(\ell-k)\bigr )+\epsilon \OO(\ell^2+k^2)~. 
\EQ(miracle)
$$
Note that the first term vanishes because $u$ is a symmetric function
and hence $u(u')^3$ is odd, and the term which is linear in $k$ and
$\ell$ vanishes as well, because of momentum conservation, so the
proof of \clm(k2k3) is complete.

\REMARK Note that a similar calculation immediately shows that the kernel
$K_3$ satisfies:
$$
|K_3(\ell,k_1,k_2)| \,\le\, C \epsilon~.
$$

We now need the following auxiliary result:
\CLAIM Lemma(4241) If $\rho_2$ and $\rho_3$ are in $\Hqr$, and if 
$\rho_1 = \rho_1(p,p')$ is a $\CC^r$ function, then
$$
\Xi(p) \,=\, \int dp'\, \rho_1(p,p') \rho_2(\dppp) \rho_3(p')
$$
is in $\Hqr$ and
$$
\| \Xi \|_{q,r}\,\le\, C \| \rho_1 \|_{\CC^r} \| \rho_2 \|_{q,r} 
\| \rho_3 \|_{q,r} ~.
$$

\PROOF Recall from Eq.\equ(x000) that $\dppp\approx p-p'$, so we are
really estimating a slightly distorted convolution. If $\dppp=(p-p')$,
the proof is easy using the definition of the norms. In the present
case, where $\dppp$ is not trivial, the result follows in a similar way by
``undoing'' part of the variable transformation which led from the
variables $\ell$, $k$ to the variables $p$, $p'$.
To simplify matters, we consider only the somewhat easier problem
of bounding
$$
\int dp' \, \Phi'(p e^{-\tau/2})  \rho_2(\dppp) \rho_3(p')~.
\EQ(a1)
$$
Using the definition of $\dppp$ this is equal to
$$
\int dp' \, \Phi'(p e^{-\tau/2})  \rho_2\bigl (e^{\tau
/2}\Phi^{-1}\bigl (
\Phi(pe^{-\tau/2})-\Phi(p'e^{-\tau/2})\bigr )\bigr )
\,\rho_3\bigl (e^{\tau
/2}\Phi^{-1}\bigl(\Phi(p'e^{-\tau/2})\bigr )\bigr )~.
\EQ(a2)
$$
Changing variables to $k=\Phi(e^{-\tau/2}p)$ and $\ell=\Phi(e^{-\tau
/2}p')$,
we get
$$
\int d\ell\,e^{\tau/2}
\rho_2 \bigl (e^{\tau /2}\Phi^{-1}(k-\ell)\bigr )\,
\rho_3 \bigl (e^{\tau /2}\Phi^{-1}(\ell)\bigr )~.
\EQ(a3)
$$
We now define a function $\Psi_\tau$ by
$$
\Psi_\tau (  e^{\tau/2} x) \,=\, e^{\tau/2}\Phi^{-1}(x)~,
$$
and note that from $\Phi(x)=x\cdot\bigl (1+\OO(x)\bigr )$ it follows that
$\Psi_\tau (y)= y\cdot \bigl (1+\OO(e^{-\tau /2}y)\bigr )$.
We can rewrite Eq.\equ(a3) as
$$
\int d\ell\,e^{\tau/2}
\rho_2 \bigl (\Psi_\tau (e^{\tau/2}(k-\ell))\bigr )\,
\rho_3 \bigl (\Psi_\tau(e^{\tau /2}\ell)\bigr )~.
\EQ(a4)
$$
We define next $\hat \rho_j(k)=\rho_j \circ \Psi_\tau$, and we see that
Eq.\equ(a4) is equal to
$$
\int d\ell\,
\hat\rho_2 \bigl (k-\ell\bigr )\,
\hat\rho_3 \bigl (\ell\bigr )~.
\EQ(a5)
$$ 
Thus, we can bound the $\Hqr$ norm of Eq.\equ(a1) by
$\|\hat\rho_2\|_{q,r}\|\hat\rho_3\|_{q,r}$, and, since $\Psi_\tau$ is
uniformly close to the identity for all $\tau$, this is in turn
bounded by $\const\|\rho_2\|_{q,r}\|\rho_3\|_{q,r}$.
This proves \clm(4241) in this special case. The extension to the
general case is easy and is left to the reader. 

We now have the necessary tools to attack the proofs of \clm(nonlin)
and \clm(nonlin2).
\LIKEREMARK{Proof of \clm(nonlin)}If we write 
out the transformation leading to $f_2$, {\it i.e.}, from
Eq.\equ(f1f2) to Eq.\equ(wequ), we get, using Eq.\equ(x0),
$$
\eqalign{
e^\tau  \cdot\bigl (f_2 (w,e^{-\tau/2 })\bigr )(p)\
 \,&=\,e^\tau 3
\chi\bigl (\Phi(pe^{-\tau /2})\bigr ) \int_{-P(\tau )}^{P(\tau )} dp'\,e^{-\tau /2}
\Phi'(p'e^{-\tau /2})\,\cr
\times&
K_2\bigl ( \Phi(pe^{-\tau /2}), \Phi(p'e^{-\tau /2})\bigr )
w(\dppp)w(p')~,\cr
}\EQ(t1)
$$
where
$$
P(\tau )\,=\,\Phi^{-1}(\HALF)e^{\tau /2}\,\approx\, \HALF e^{\tau
/2}~.
$$
We bound $|K_2\bigl ( \Phi(pe^{-\tau /2}), \Phi(p'e^{-\tau /2})\bigr
)|$ by
$C \epsilon |\Phi(pe^{-\tau /2})^2 + \Phi(p'e^{-\tau /2})^2|$
using \clm(k2k3).
Since the expressions $\Phi(pe^{-\tau /2}),$ and $\Phi(p'e^{-\tau /2}),$ in
Eq.\equ(t1) are bounded, and $\Phi(x)=x(1+\OO(x))$,
we can extract another factor of $e^{-\tau
/2}$ and get a bound on
$e^{\tau }f_2$ of the form
$$
\eqalign{
\const e^{\tau /2}&\chi\bigl (\Phi(pe^{-\tau /2})\bigr )
\int _{-P(\tau )}^{P(\tau )} dp'\,  \bigl (\bigl|\Phi(pe^{-\tau
/2})\bigr |+ \bigl |\Phi(p'e^{-\tau /2})\bigr | \bigr )
\cdot|w(\dppp)w(p')|\cr
\,&\le\,\const \chi\bigl (\Phi(pe^{-\tau /2})\bigr )
\int _{-\infty }^\infty  dp'\,
|p+p'| |w(\dppp)w(p')|~.\cr
}
\EQ(t111)
$$
If $w$ is in $\Hqr$, then with the aid of \clm(4241), we can estimate the $H_{q-1,r}$
norm of Eq.\equ(t1) by $C \| w \|_{q,r}^2$.
Note further, that from the above discussion it is also clear
that $e^{\tau} f_2(w^\c,e^{-\tau/2})(p)$ is also a smooth function
of $e^{-\tau/2}$.

\REMARK The factors $|p|, |p'|$ are responsible for the loss of one power in
the norm estimate of \clm(nonlin). It is only in the study of the flow
within the invariant manifold that we will need the second order bound of
\clm(k2k3). 

\REMARK Note that the nonlinear terms depend (implicitly) on the constant
$t_0$ which entered the definition of the new temporal variable $\tau$.
However, all the estimates above (as well as those which follow
in the proof of \clm(nonlin2)) are independent of this constant.

The bound on $f_3$ is similar, but no additional regularization is
needed, since there are {\em two} integrations, each of which
contributes a factor $e^{-\tau /2}$. We leave this to the reader.
The proof of the asserted bounds of  Eq.\equ(fbound2) is complete.

We now turn to the estimates of the nonlinear terms $f_4$ and $g$.
Because these terms involve the $w^\s$, we begin with
a discussion of the appropriate function space for these components.
These were defined in Section 3, but we repeat
them here for convenience.
Recall that $w^\c \in \Hqr$, while $w^\s \in \Hqr \oplus
\Hqrn $, where
$$
\eqalign{
\Hqrn  \,=\,\{ w = w(p;x) & ~|~ w(p;x) = w(p;x+2 \pi), \cr
& (1 - \partial_x^2)^{\nu/2}  (1-\partial_p^2)^{r/2}(1+p^2)^{q/2} w 
\in L^2(\real \times [-\pi, \pi]) \} ~. 
\cr}
$$
The fact that $w^\s$ is an element of the direct sum of two spaces
reflects the fact (see the paragraph preceding \equ(system),
and then \equ(system2) ) that it has two components, the
first of which comes from the central branch of the spectrum
of $L_{\ell}$, but with $\ell$ localized away from zero, and the
second component coming from the stable branches of the spectrum
of $L_{\ell}$. In a slight abuse of notation we will denote
by $\| w^\s \|_{\HHqrn}$ the sum of the $\Hqr$ norm of the 
first component of $w^\s$ and the $\Hqrn $ norm of the second component,
and by $\| w^\s\|_{q,r,\nu}$, we will mean the $\Hqrn $ norm of
just the second component.
\REMARK An easy fact which will be useful later is that if
we expand $w(p;x) \in \Hqrn $ in a Fourier series with respect to $x$,
$$
w(p;x) \,=\,\sum_{n=-\infty}^{\infty} e^{inx} \hat{w}_n(p) ~,
$$
then the $\Hqrn $ norm of $w$ is equivalent to the norm
$$
\| w \|_{{H}_{q,r,\nu}}^2 \,=\,\sum_{n = -\infty}^{\infty}
(1 + n^2 )^{\nu} \| \hat{w}_n \|_{q,r}^2 ~.
\EQ(521)
$$
Thus we will use the two norms interchangeably.

Now consider
$$
e^\tau f_4(w^\c,w^\s e^{-\tau/2} , e^{-\tau/2})~.
\EQ(f4)
$$
We shall concentrate on the most ``dangerous'' piece which is the
quadratic term with one factor of $w^\c$ and one of $w^\s$. Other terms
are ``less dangerous'' in the sense that they contain either more
factors of $w^\s$ each of which contributes a small factor of
$e^{-\tau /2}$, or more convolutions which again contribute a factor
of $e^{-\tau /2}$. The quadratic piece of \equ(f4) has the form
$$
\eqalign{
e^\tau 3 \chi&\bigl( \Phi (p e^{-\tau/2})\bigr )
\int dx\, \bar \phi _{ \Phi (p e^{-\tau/2})}(x)\, u_\epsilon (x)\,\cr
\times&\int_{-P(\tau)}^{P(\tau)}dp'\, e^{-\tau /2}\Phi'(p'e^{-\tau/2 })
w^\c\bigl(\dppp\bigr )
\cr
&~~\times \phi _{\gppp}(x)\,
e^{-\tau/2} w^\s(p';x)~.
\cr
}\EQ(f22)$$ 

As we mentioned above, $w^\s$ has two components---one in $\Hqr$, and one
in $\Hqrn $. The contribution from the component in $\Hqr$ is bounded
by the same techniques used to control $f_3$---note that it
is not necessary to extract any additional factors of $e^{-\tau/2}$,
since we get one from the integration, and one from the fact that
each factor of $w^\s$ is multiplied by $e^{-\tau/2}$. Thus, we restrict
our attention to the component of $w^\s$ in $\Hqrn $, which is where
the new ingredients are necessary.

Interchanging the order of the $x$ and $p'$ integrals, we use
\clm(4241), with 
$$
\eqalign{
\rho_1(p,p')\,&=\,\sup_x\,\bigl| 3 \chi\bigl( \Phi (p e^{-\tau/2})\bigr
)\Phi'(p'e^{-\tau/2 })\,\phi _{ \Phi (p
e^{-\tau/2})}(x)\, \phi _{\gppp}(x)u_\epsilon (x)\bigr |~,\cr
\rho_2(r)\,&=\,|w^\c(r )|~,\cr
\rho_3(p')\,&=\,\bigl|\int dx\, w^\s(p';x)\bigr |~.\cr
}
$$
Since $\phi_\ell(x)$ and $u_\epsilon (x)$ are smooth, $2\pi$-periodic
functions of $x$, and $\| \rho_1 \|_{\CC^r}$ is bounded,
the \clm(4241) implies that the $\Hqr$ norm
of \equ(f22) is bounded by
$$
C \| w^\c \|_{q,r} \,\| \int dx\,  
 w^\s(\cdot;x) \|_{q,r}~.
\EQ(f22a)
$$
The $\Hqr$ norm of the integral can be bounded by
$$
\sup_x \| w^\s(\cdot;x) \|_{q,r}\,\le\, C \| w^\s \|_{\Hqrn }~,
\EQ(f22b)
$$
provided $\nu > 1/2$, where we used Sobolev's inequality to estimate the
supremum over $x$.
Inserting \equ(f22b) into \equ(f22a) yields the bound claimed in 
\equ(fboundf4).

The remaining terms in $f_4$ can be bounded in a similar fashion, but
as noted above, they will tend to $0$ as $\tau \to \infty$. In fact,
they will be bounded by $C \epsilon e^{-\tau/2}$.

\LIKEREMARK{Proof of Eq.\equ(fboundg) of \clm(nonlin2)}We 
finally bound the non-linear term
$$
e^{\tau/2}g(w^\c,w^\s e^{-\tau /2},e^{-\tau/2 })~.
\EQ(g1)
$$

In bounding $e^{\tau/2} g(w^\c,w^\s e^{-\tau/2}; e^{-\tau/2})$, recall that
just as $w^\s$ did, this expression will have two components---one
in $\Hqr$, and one in $\Hqrn $. The component in $\Hqr$ is
bounded using exactly the same techniques used to
control the  term $f_4$ above, so we concentrate
here on explaining the new ingredients necessary to bound the
component in $\Hqrn$.

As in the bound on $f_4$, the potentially largest terms are
those of minimal order, because each additional order provides a
factor of $e^{-\tau /2}$. So we look at the terms which are
quadratic and which are of order $w^\c w^\c$, $w^\c w^\s$, and $w^\s w^\s$,
respectively.
The first term leads us to study
$$
e^{\tau/2}P_p^\perp \biggl (
u_\epsilon (x)\,  \int_{-1/2}^{1/2} dp'\,
V^\c(p-p') V^\c(p') \phi_{p'}(x)\, \phi_{p-p'}(x)
\biggr )~.
\EQ(g10)
$$
Rescaling as in \equ(system2), we see we must bound
$$
\eqalign{
P^\perp_{\Phi (p e^{-\tau/2})}\biggl( u_{\epsilon}(x)\, & 
\int_{-P(\tau)}^{P(\tau)} d{p'} \,\Phi'({p'}e^{-\tau/2}) 
\cr & \times\phi_{\Phi ({p'} e^{-\tau/2})}(x)\, \phi _\gppp(x)\, w^\c({p'})
\,w^\c(\dppp)\biggr) ~.
\cr}
\EQ(g15)
$$
Note that the prefactor of $e^{\tau/2}$ has disappeared due
to the factor of $e^{-\tau/2}$ which we gain as usual from the
change of variables.

Since the projection $P^\perp_{\ell}$ has bounded norm and is a smooth
function of $\ell$, we can discard this factor at the price of 
introducing an overall constant in the estimate.
Note next that the square of the $\Hqrn $ norm of the remaining expression is
equal to:
$$
\eqalign{
\Bigl\| \int_{-P(\tau)}^{P(\tau)} & d{p'}\, \Phi'({p'}e^{-\tau/2})
 w^\c({p'}) w^\c(\dppp) \cr &\times \|  u_{\epsilon}(x)\, 
\phi_{\Phi ({p'} e^{-\tau/2})}(x)\, \phi _\gppp(x)\, \|_{H^{\nu}(dx)}^2 \Bigr
\|_{\Hqr(dp)}^2~,
\cr}
\EQ(g20)
$$
where the $H^{\nu}$ norm is the $H^{\nu}$-Sobolev norm of the quantity
$$
u_{\epsilon}(x)\,
\phi_{\Phi ({p'} e^{-\tau/2})}(x)\, \phi _\gppp(x)~,
$$ 
considered as a function of $x$,
and the $H_{q,r}$ norm is the norm of the resulting function of $p$.
Since $u_{\epsilon}(x)\,
\phi_{\Phi ({p'} e^{-\tau/2})}(x)\, \phi _\gppp(x)$ is a smooth function of
$x$, ${p'}$, and $p$, there exists a smooth, bounded function 
$\psi(p,{p'})$, such that 
$$
\psi(p,{p'}) \,=\, \| u_{\epsilon}(x)\,
\phi_{\Phi ({p'} e^{-\tau/2})}(x)\, \phi _\gppp(x)\, \|_{H^\nu(dx)}~.
\EQ(psidef)
$$
But now, by \clm(4241), we can conclude that \equ(g20) is bounded by
$$
\|  \int_{-P(\tau)}^{P(\tau)} d{p'}\, \Phi'({p'}e^{-\tau/2})
 w^\c({p'}) w^\c(\dppp) \psi(p,{p'}) \|_{\Hqr(dp)}^2 \,\le\,
C \| \Phi' \psi \|_{\CC^r}^2 \| w^\c \|_{q,r}^4~.
\EQ(g23)
$$

We next consider the quadratic term in $g$ which contains one factor
of $w^\c$ and one factor of $w^\s$. In this case, the analog of \equ(g15)
is
$$
\eqalign{
e^{-\tau/2} P^\perp_{\Phi (p e^{-\tau/2})}\biggl( & u_{\epsilon}(x) 
\int_{-P(\tau)}^{P(\tau)} d{p'}\, \Phi'({p'}e^{-\tau/2}) 
 \cr
& \times
 \phi _\gppp(x)\, w^\c(\dppp) w^\s({p'};x)\biggr )~.
\cr}
\EQ(g25)
$$
Note that in this case, we pick up an extra factor of $e^{-\tau/2}$,
in comparison with \equ(g15), since each factor of $w^\s$ is multiplied
by this exponential.

Once again, we must contend with the fact that $w^\s$ has two components.
However, the component in $H_{q,r}$ behaves exactly as in the estimates
leading to \equ(fbound2), so we concentrate on the component in $\Hqrn $.

As above, the projection operator can be dropped at the cost of
an overall constant, and we are left with the task of bounding the
$\Hqrn $ norm of the remainder. The square of this norm is equal to
$$
\eqalign{
\Bigl\| \int_{-P(\tau)}^{P(\tau)} & d{p'}\, \Phi'({p'}e^{-\tau/2}) 
w^\c(\dppp)\cr & \times\|  u_{\epsilon}(x)\, 
\phi_{\Phi ({p'} e^{-\tau/2})}(x)\, \phi _\gppp(x)\, w^\s({p'};x)
\|_{H^\nu(dx)}^2 \Bigr\|_{H_{q,r}(dp)}^2  \cr
& \,\le\,  C \| \Phi' \|_{\CC^r}\, \| w^\c  \|_{H_{q,r}}^2 \cr
& \quad \times\Bigl\|  \|u_{\epsilon}(x)\,
\phi_{\Phi ({p'}  e^{-\tau/2})}(x)\,\phi _\gppp(x)\, w^\s({p'};x) 
\|_{H^\nu(dx)}^2 \Bigr\|_{H_{q,r}(dp')}^2~,
\cr}
\EQ(g40)
$$
by \clm(4241). Note that the pair of norms on the last factor is
equivalent to computing the square of the $\Hqrn $ norm of
$$
u_{\epsilon}(x)\,
\phi_{\Phi ({p'}  e^{-\tau/2})}(x)\,\phi _\gppp(x)\, w^\s({p'};x)~.
\EQ(g45)
$$
Since $u_{\epsilon}$, $\Phi$, $\phi_{\Gamma}$, and $\Delta$ are
all smooth, bounded functions, we see just by writing out the
definition of the norm that this is bounded by
$$
C \|w^\s\|_{\Hqrn }^2~.
\EQ(g47)
$$
If we estimate the term quadratic in $w^\s$ in a similar fashion,
and combine this estimate with that in 
\equ(g23) we see that the quadratic
terms in $e^{-\tau/2} g(w^\c, w^\s e^{-\tau/2}; e^{-\tau/2})$ are bounded
in $\HHqrn $, by
$$
C ( \| w^\c \|_{H_{q,r}} + e^{-\tau/2} \|w^\s\|_{\HHqrn } )^2~.
\EQ(g50)
$$

Analogous estimates of the cubic terms lead to a bound 
$$
C e^{-\tau/2} ( \| w^\c \|_{\Hqr}  + e^{-\tau/2} \|w^\s\|_{\HHqrn } )^3~,
\EQ(g60)
$$
where the additional factor of $e^{-\tau/2}$ comes from the additional
convolution. Combining \equ(g50) and \equ(g60)
leads to the estimate in \equ(fboundg) and completes the proof of 
\clm(nonlin2).
\def\actualnumber{B}
\SECTION Bounds on the linear operators

In this Appendix, we give bounds on the semi-group generated by $L$
and on the linear evolution defined by $M_{\exp(-\tau/2)}$.
\SUBSECTION Bound on the semi-group generated by $L$

We consider the semi-group whose generator is 
$L=\partial_x^2+\HALF x\partial_x+\HALF$.
Note that in this section, for ease of use, we define $L$ in the
Fourier transformed variables, compared to Section 2.
Fourier transformation is an isomorphism from $H_{q,r}$ (in the
$p$-variables) to $H_{r,q}$ (in the $x$-variables), so establishing
estimates on the semigroup associated to $\partial_x^2+\HALF x\partial_x+\HALF$
in the space $H_{r,q}(dx)$ will immediately imply estimates
on the representation of $L$ in the $p$-variables in the space
$H_{q,r}(dp)$. In order to avoid confusion, in what follows we
will denote by $| \cdot |_{q,r}$ the norm on $H_{r,q}(dx)$.
With this notation, the norms $\|\cdot\|_{q,r}$ and $|\cdot|_{q,r}$
resp.~the spaces
$H_{q,r}(dp)$ and $H_{r,q}(dx)$ are equivalent.

The integral kernel of the semigroup generated by $L$ is given by [GJ]
$$\bigl (e^{\tau L}v)(x)\,=\,{1\over \sqrt{ 4\pi a(\tau )} }
\int dz\, 
e^{-z^2/( 4a(\tau ))} v(e^{\tau /2}(x+z))
~,
$$
where $a(\tau )=1-e^{-\tau }$.
If we denote by $T$ the operator of multiplication by $\exp(x^2/8)$
and by $H_0$ the harmonic oscillator Hamiltonian
$H_0=\partial_x^2-x^2/16+1/4$, (note the unconventional sign!), then
$$
L\,=\,T^{-1} H_0 T~.
$$
Thus, the two operators $L$ and $H_0$ are ``the same,'' but they act on two
quite different spaces.
If the $\{\phi_j\}_{j\ge0} $ are the eigenfunctions 
of $H_0$, then the $\psi_j=T^{-1}\phi_j$ are the eigenfunctions of $L$,
with the same eigenvalues $\mu_j=-{j/ 2}$.
We let $P_nf = \sum_{j\le n} \psi_j(\psi_j, f)_q$,
where $(\cdot,\cdot)_q$ is the scalar product 
$$
\eqalign{
(f,g)_q\,&=\,(Tf,T (1-L_0)^q g)\,=\,(Tf,(1-H_0)^q Tg)~.\cr
}
$$

We next show that for $n<q-2$, the operator $P_n$
is bounded in $H_{r,q}(dx)$. First of all, the eigenfunctions
$\phi_j $ are bounded by $\OO(1)|x|^j e^{-x^2/8}$ at large $x$.
Therefore, we also have 
$\psi_j= T^{-1} \phi_j \in H_{r,q}(dx)$, since it decays
exponentially. Finally, 
$$
( \psi_j, f)_q\,=\,(T\psi_j,(1-H_0)^q T f)\,=\,|1-\mu_j|^q(\phi_j,Tf)~,
$$
and the last scalar product is bounded if $f\in H_{r,q}(dx)$ when $r>j+2$,
since, with a weight function $W(x)=(1+x^2)^{1/2}$,
$$
\eqalign{
|(\phi_j,Tf)|\,&\le\,C \bigl |(W^j ,f)\bigr | \,\le\, C\bigl |(W^{-1},
W^{j+1}f)\bigr |\cr
\,&\le\, C \|W^{j+1} f\|_2\,\le\,C |f |_{0,r}~.
\cr
}
$$
Thus $P_n$ is defined. We let $Q_n=1-P_n$ (in $H_{r,q}(dx)$).
\CLAIM Theorem(semigroupa) For every $\epsilon >0$, 
there are an $m_0$ and a function $r(m,q)$
such that for every
$m\ge m_0 $, every $q\ge1$ 
and every $r\ge r(m,q)$, there
is a $C=C(q,r,m)<\infty $ such that
$$
| e^{\tau  L} Q_m v |_{q,r}\,\le\,
{C(q,r,m)\over \sqrt{a(\tau )}} e^{- \tau |\mu _{m}|+\tau \epsilon } 
| v |_{q-1,r}~.
\EQ(decay)
$$

\REMARK The function $r(m,q)$ is of order $\OO(m+q)$.
\PROOF To explain the strategy of the proof, we need some
notation. Let $P_n^{(0)}$ denote the projection in $H_{0,q}(dx)$ onto the
subspace spanned by $\{\phi_j\}_{j\le n}$ and let
$Q_n^{(0)}=1-P_n^{(0)}$.
Then, formally, $TQ_n = Q_n^{(0)} T$, and $LQ_n= T^{-1} H_0 Q_n^{(0)}
T$.
This suggests that $L $ restricted to $Q_n$
has no spectrum in the half-plane
$\{ z ~|~ {\rm Re\,} z > -|\mu_{n+1}| \}$, and
thus one can understand
the decay in Eq.\equ(decay). The square-root singularity at $\tau =0$ is
related to our gain in smoothness.
The problem is that $TQ_n =Q_n^{(0)}T$ is
ill-defined. However, it will be well defined if we localize near
$x=0$. In that region, the heuristic argument will be seen to be
valid, whereas in the complement of such a region, when $|x|>R$, decay
will be shown by direct methods,
using the explicit form of the integral kernel.

We study first the quantity $ \chi_Re^{\tau L}$, where $\chi_R$ is a
smooth characteristic function which vanishes for $|x|<R$ and is equal
to 1 for $|x|>4R/3$. Thus we study a region far from the origin.
Our bound is
\CLAIM Proposition(outer) For every $q\ge1$ and every $r\ge0$
there exists a $C(q,r)<\infty $ such that
for all $v\in H_{r,q}(dx)$ one has
$$
\eqalignno{
| \chi_R   e^{\tau L} v|_{q,r} \,&\le\,
{C(q,r)\over \sqrt{a(\tau )}} e^{\tau q/2} \left ( e^{-\tau r/2} +
e^{-3R^2/16}\right ) |v |_{q-1,r}~,\NR(outer)
|  \chi_Re^{\tau L} v|_{q,r} \,&\le\,
{C(q,r)} e^{\tau q/2} \left ( e^{-\tau r/2} +
e^{-3R^2/16}\right ) |v |_{q,r}~.\NR(outer1)
}
$$

\CLAIM Corollary(outer0) For every $q\ge1$ and every $r\ge0$
there exists a $C(q,r)<\infty $ such that
for all $v\in H_{r,q}(dx)$ one has
$$
\eqalignno{
|e^{\tau L} v|_{q,r} \,&\le\,
{C(q,r)\over \sqrt{a(\tau )}} e^{\tau q/2}  |v|_{q-1,r}~,
\NR(outer0)
|e^{\tau L} v|_{q,r} \,&\le\,
{C(q,r)} e^{\tau q/2}  |v|_{q,r}~.
\NR(outer01)
}
$$

\LIKEREMARK{Remarks}The 
improvement over [W] is that we ``gain'' a derivative
in $x$. The corollary follows easily by repeating the proof of
\clm(outer) with $R=0$.

\PROOF We let $\D=\partial_x$ and denote, as before, by $W$ the operator
of multiplication by $(1+x^2)^{1/2}$.
Then
$$
|  \chi_R e^ {\tau L} w |_\qr^2 \quad {\rm and}\quad
\sum_{q'\le q}
 \|W^r\D^{q'} \chi_R e^{\tau L }  w \|_2^2
$$
are equivalent.
We shall only consider the term with the highest derivative,
because only there is the issue of regularization important. Thus we
are led
to bound
$$
X^2\,=\,
\| W^r \D^q \chi_R e^{\tau L }  w\|_2^2
~.
$$
Since $L=\partial_x^2 +\HALF x\partial_x +\HALF$, 
a quick calculation shows that
$$
\D^q e^{\tau L} \,=\, e^{\tau q/2} e^{\tau L}\D^q~.
$$
The diverging factor $\exp(\tau q/2)$ will appear in the final bound.
Note now that
$$
\left (e^{\tau L}\D^q v\right  )(x)\,=\,
{1\over\sqrt{4\pi a(\tau )}}e^{\tau /2}
\int dz\,
e^{-z^2/( 4a(\tau ))} \bigl(\D^q v\bigr)(e^{\tau /2} (x+z))~,
\EQ(s1)
$$
which upon integrating by parts becomes
$$
{1\over\sqrt{4 \pi a(\tau )}}
\int dz\,
{z\over 2 a(\tau )}e^{-z^2/( 4a(\tau ))} \bigl(\D^{q-1}
 v\bigr)(e^{\tau /2} (x+z))~.
$$
Use now the Schwarz inequality in the form (for positive $f$ and $g$),
$$
\eqalign{
\|f * g\|_2^2\,&=\,\int dx\,\int dz_1\,dz_2\, f(z_1)f(z_2) g(x-z_1)
g(x-z_2)\cr
\,&\le\,\int dz_1\,dz_2\, f(z_1) f(z_2) \|g(\cdot-z_1)\|_2
\|g(\cdot-z_2)\|_2
\cr
\,&=\,\left (\int dz\, f(z) \|g(\cdot-z)\|_2\right )^2~.\cr
}
$$
This leads to a bound 
$$
\eqalign{
X\,&\le\,
    {e^{\tau q/2}\over \sqrt{4\pi a(\tau )}}
\int_{R_1\cup  R_2} \kern -1em dz\,{|z|\over 2 a(\tau )}e^{-z^2/( 4a(\tau ))}
\|W^r \chi_R\bigl(\D^{q-1}
  w\bigr)(e^{\tau /2} (\cdot+z))\|_2\cr
\,&\equiv\, X_1 +X_2~,\cr
}
\EQ(x1)
$$
where we let $R_1=\{x~:~|x|<7R/8\}$ and $R_2=\real\setminus R_1$.
To be more precise, we define $\chi_R$ by the scaling of a fixed
function:
$\chi_R(x)=\chi(x/R)$. If $R\to \infty $, then $\partial_x
\chi_R(x)=\OO(R^{-1})$ and therefore it is uniformly bounded. 
\CLAIM Lemma(52) (Lemma A.2 of [W]). One has the bounds
$$
\|W^r\chi_R(\cdot)
  v(e^{\tau /2} (\cdot+z))\|_2^2\,\le\,\cases
{ C e^{-r\hide{(-r-{1\over 2})}\tau } |v|_{0,r}^2~,&if $ |z| \le 7R/8$,\cr
C \hide{e^{-{1\over 2}\tau }}(1+z^2)^r |v|_{0,r}^2~,&if $ |z| >7R/8$.\cr
}
\EQ(both)
$$

\LIKEREMARK{Proof of \clm(52)}Consider 
first the case $|z|\le 7R/8$. Since $|x|>R$ on the support of $\chi_R$, we
have $|x+z|\ge |x|/8$ and hence 
$$(1+x^2)/\bigl (1+(e^{\tau /2} |x+z|)^2\bigr
)\,\le\, \const e^{-\tau }~.
$$
Using this, we bound
$$
\eqalign{
\int_{R_1} dx\,& (1+x^2)^{r} |\chi_R(x)v(e^{\tau /2}(x+z))|^2\cr
\,&=\,
\int_{R_1} dx\, {(1+x^2)^r\over \bigl (1+(e^{\tau /2} |x+z|)^2\bigr
)^r }\cdot \bigl (1+(e^{\tau /2} |x+z|)^2\bigr )^r|v(e^{\tau /2}(x+z))|^2\cr
\,&\le\,\const e^{-\tau r} e^{-\tau /2} |v|_{0,r}^2\,\le\,\const e^{-\tau r} |v|_{0,r}^2~.
}
$$
In the second case, we get
$$
\eqalign{
\int_{R_2} dx\,& (1+x^2)^{r} |\chi_R(x)v(e^{\tau /2}(x+z))|^2\cr
\,&=\,
e^{-\tau /2}\int dy\, {\bigl (1+(e^{-\tau /2} y-z)^2\bigr )^r\over 
(1+y^2)^r} (1+y^2)^r|v(y)|^2\cr
\,&\le\,\const e^{-\tau /2} (1+z^2)^r |v|_{0,r}^2
\,\le\,\const  (1+z^2)^r |v|_{0,r}^2~.
}
$$
The proof of \clm(52) is complete.

Continuing the proof of \clm(outer),
we first bound the integral over $R_1$ in Eq.\equ(x1).
We get from the first alternative of \clm(52),
$$
\eqalign{
X_1\,&=\,
{1\over \sqrt{4\pi a(\tau )}}e^{\tau q/2}\int _{R_1} dz\,
{|z|\over \sqrt{2 a(\tau )}}e^{-z^2/( 4a(\tau ))}
\|W^r\chi_R\bigl(\D^{q-1}
  w\bigr)(e^{-\tau /2} (\cdot+z))\|_2\cr
\,&\le\,
\const {1\over \sqrt{4\pi a(\tau )}}e^{\tau q/2}\int _{R_1} dz\,
{|z|\over \sqrt{2 a(\tau )}}e^{-z^2/( 4a(\tau ))}
e^{-\tau r/2} |w |_{q-1,r}
\cr
\,&\le\,
\const{1\over \sqrt{4\pi a(\tau )}} e^{\tau (q/2-r/2)} |w|_{q-1,r}~.}
$$
Similarly, using the second alternative in Eq.\equ(both), we get
$$
\eqalign{
X_2\,&=\,
{1\over \sqrt{4\pi a(\tau )}}e^{\tau q/2}\int _{R_2} dz\,
{|z|\over \sqrt{2 a(\tau )}}e^{-z^2/( 4a(\tau ))}
\|W^r\chi_R\bigl(\D^{q-1}
  w\bigr)(e^{\tau /2} (\cdot+z))\|_2\cr
\,&\le\,
\const {1\over \sqrt{4\pi a(\tau )}}e^{\tau q/2}\int _{R_2} dz\,
(1+z^2)^{r/2}{|z|\over \sqrt{2 a(\tau )}}e^{-z^2/( 4a(\tau ))}
|w|_{q-1,r}~.
\cr
\,&\le\,
\const {1\over \sqrt{4\pi a(\tau )}}e^{\tau q/2\hide{-\tau /4}} 
e^{-3R^2/16} |w |_{q-1,r}~,
}
$$
since $3/16< (7/8)^2/4$.
Note that the constants above depend on $r$ and $q$, but can be chosen
uniformly for all $R\ge 1$. 
The proof of Eq.\equ(outer) is complete.
Omitting the integration by parts in Eq.\equ(s1), the assertion
Eq.\equ(outer1) follows in the same way. The proof of \clm(outer) is complete.

We next study $e^{\tau L} Q_n(1-\chi_R) w$. We have the following bound
\CLAIM Proposition(innerm) For every $\epsilon >0$, $q\ge1$, and every
$r\ge0$ there is a  
$C(\epsilon ,q,r)<\infty $ such
that 
$$
|e^{\tau L} Q_n(1-\chi_R) w |_{q,r}\,\le\,{C(\epsilon ,q,r)
\over \sqrt{a(\tau )}}
e^{-|\mu _{n+1}|\tau +\tau \epsilon } e^{R^2/6} |w|_{q-1,r}~.
\EQ(innerm)
$$

\PROOF Recall that $T=e^{x^2/8}$ and that $L=T^{-1} H_0 T$.
The operator $T(1-\chi_R)$ is bounded and $\|T(1-\chi_R)\|\le
\const e^{R^2/6}$.
Therefore we have
$$
\eqalign{
Q_n T (1-\chi_R)\,&=\, (1-P_n)T (1-\chi_R) \,=\,
T(1-\chi_R) - TP_n^{(0)} (1-\chi_R)\cr
\,&=\,T( 1-P_n^{(0)})(1-\chi_R) \,=\,TQ_n^{(0)} (1-\chi_R)~,\cr
}
$$ 
where $Q_n^{(0)}$ is the
orthogonal projection onto the complement of the subspace spanned by
the first $n$ eigenvalues of $H_0$ in $H_{q,0}$.
It is easy to see that on $H_{r,q}(dx)$, the operator $(1+x^2)^{1/2}
(1-H_0)^{-1/2} $ is bounded. Thus, we get, using the spectral
properties of $H_0$ (on $Q_n^{(0)}$),
$$
\eqalign{
|e^{\tau H_0} T Q_n (1-\chi_R) w|_{q,r} \,&=\,
\tau ^{-1/2}\cr
&\times~|(1-H_0)^{-1/2}e^{\tau H_0} (\tau (1-H_0))^{1/2} Q_n^{(0)} T(1-\chi_R)
w|_{q,r} \cr
\,&\le\,\const  \tau ^{-1/2} | e^{\tau H_0} (\tau (1-H_0))^{1/2} Q_n^{(0)} T(1-\chi_R)
w |_{q-1,r} \cr
\,&\le\,\const \tau ^{-1/2} e^{-\tau |\mu_{n+1}|+\tau \epsilon } 
|T(1-\chi_R)w |_{q-1,r}\cr
\,&\le\,\const \tau ^{-1/2}e^{-\tau |\mu_{n+1}|+\tau \epsilon } e^{R^2/6} |w|_{q-1,r}~.
}\EQ(h0)
$$
The proof of \clm(innerm) is complete.

\LIKEREMARK{End of proof of \clm(semigroupa)}We first rewrite $e^{\tau L}
Q_n$ as
$$
e^{\tau L}Q_n \,=\,e^{\tau L/2}Q_ne^{\tau L/2} 
\,=\,e^{\tau L/2}Q_n\chi_Re^{\tau L/2}  + 
e^{\tau L/2}Q_n(1-\chi_R)e^{\tau L/2} ~.
$$
The second term can be bounded by \clm(innerm) and Eq.\equ(outer01)
as
$$
\eqalign{
|e^{\tau L/2}Q_n(1-\chi_R)e^{\tau L/2}  w|_{q,r}\,&\le\,
{C\over \sqrt{a(\tau )}} e^{R^2/6 - \tau |\mu _{n+1}|/4} 
|e^{\tau L/2}  w |_{q-1,r}\cr
\,&\le\,
{C\over \sqrt{a(\tau )}} e^{R^2/6 - \tau |\mu
_{n+1}|/4}e^{\tau q/4\hide{(q/4-1/8)}} |w |_{q-1,r}~. \cr
}
$$
This quantity is bounded by
$$
{C\over \sqrt{a(\tau )}} e^{ - \tau n/8} |w|_{q-1,r}~,
\EQ(f1)
$$
provided $n$ is much larger than $q$ and $R^2/6< \tau n/16$.
The first term can be bounded by Eq.\equ(outer01) and Eq.\equ(outer)
as
$$
\eqalign{
|e^{\tau L/2}Q_n\chi_Re^{\tau L/2}  w|_{q,r}\,&\le\,
{C\over \sqrt{a(\tau )}} e^{\tau q/4{\hide{(q/4-1/8)}}}
 |\chi_Re^{\tau L/2}  w |_{q-1,r}\cr
\,&\le\,
{C\over \sqrt{a(\tau )}} e^{\tau q/2\hide{(q/2-1/4)}}\left (e^{-\tau r/2} +
e^{-3R^2/16}\right ) |w|_{q-1,r}\cr
\,&\le\,{C\over \sqrt{a(\tau )}}e^{-\tau n/8} |w|_{q-1,r}~,\cr
}
\EQ(f2)
$$
provided $r\ge n/4+q$ and $3 R^2/16 \ge \tau ( n/8+q/2)$. Note that the
conditions on $R$ from the first and second term are compatible

Combining Eqs.\equ(f1)--\equ(f2), we get
$$
| e^{\tau L} Q_n w |_{q,r}\,\le\,{C\over \sqrt{a(\tau )}} e^{-\tau n/8}
 |w|_{q-1,r}~.
\EQ(ff4)
$$ 
It remains to improve the decay rate from $n/8$ to $|\mu _{m+1}|$.
The idea is to just take $n=8(m+1)$. Then we find
$$
e^{\tau L}Q_m\,=\,e^{\tau L}Q_n Q_m + e^{\tau L} P_m Q_m + e^{\tau L} (P_n-P_m) Q_m~.
\EQ(trick)
$$
The first term is bounded by Eq.\equ(ff4), and $m/8>-|\mu _{n+1}|$.
The second term vanishes and the third is diagonalized explicitly:
$$
e^{\tau L} (P_n-P_m)Q_m\,=\,T^{-1} e^{-\tau H_0} T (P_n-P_m) Q_m\,=\,
T^{-1} e^{-\tau H_0} (P_n^{(0)}-P_m^{(0)})T Q_m~.
$$ 
We are operating here on the finite dimensional subspace spanned by
the eigenvectors $\phi_{m+1}$, $\dots,\phi_n$, and there the technique of
Eq.\equ(h0) yields a bound
$$
{C\over \sqrt{a(\tau )}} \sqrt{\tau |\mu _{m+1}|}e^{-|\mu _{m+1}| \tau }~.
$$
Combining this with the bound on the first term in 
Eq.\equ(trick), we complete the proof of \clm(semigroupa).
\SUBSECTION The linear evolution generated by $M_{\eta ,2}$

In this section, we deal with the problem of giving bounds on
the linear evolution
generated by the operator $M_{\eta ,2}$, which is defined by
$$
M_{\eta ,2}\,=\,  M_{\eta,2,0}  \,\oplus \,
\mathop\oplus_{n=2}^\infty  M_{\eta ,2,n}~,
$$
where 
$$
M_{\eta ,2,n}\,=\,\biggl(\bigl (\epsilon ^2-
(1+(in+i\Phi(p\eta))^2\bigr )^2
-\KK\bigl (\Phi(p\eta)\bigr )\biggr)-\eta^2\HALF
p\partial_p~.
$$
We want to bound the solution $U_{n,\tau} $ of the equation
$$
e^{-\tau }\partial_\tau  U_{n,\tau }\,=\, M_{\exp(-\tau /2),2,n} U_{n,\tau }~,
\EQ(mmm)
$$
with $U_{n,0}=1$.
Recall the definition of $L=-p^2-\HALF p\partial_p$,
and rewrite $M_{\exp(-\tau/2),2,n} $ as
$$\eqalign{
M_{\exp(-\tau/2),2,n} \,&=\, \biggl(\epsilon^2 - \bigl (1+(in +
i\Phi(pe^{-\tau/2}))^2\bigr )^2 
	-\KK(\Phi(p e^{-\tau/2}))\biggr)  - e^{-\tau} \HALF p \partial_p \cr
\,&=\, \biggl(\epsilon^2 - \bigl (1+(in +
i\Phi(pe^{-\tau/2}))^2\bigr )^2 
	-\KK(\Phi(p e^{-\tau/2})) \biggr )+ e^{-\tau} p^2 +e^{-\tau} L \cr
\,&=\, X_n(p e^{-\tau/2})+e^{-\tau} L ~, \cr}
$$
where $X_n(\xi)= \epsilon^2 - \bigl (1+(in + i\Phi(\xi))^2\bigr )^2 
	-\KK(\Phi(\xi))  + \xi^2 $.
We want to solve Eq.\equ(mmm):
$$
e^{-\tau} \partial_{\tau} U_{n,\tau} \,=\,
(e^{-\tau} L + X_n(p e^{-\tau/2}) ) U_{n,\tau} ~,
$$
with initial condition $U_{n,0} =  1$.
Observe now that $X_n$ is an operator of multiplication by a function of
$p\eta$.
Since the commutator $[p^m,-p^2 -\HALF p\partial_p]$ is equal to
${m\over 2}p^m$, we find $[h(p),L] = \HALF p h'(p)$,
and, furthermore, 
$$
e^{h(p)} L \,=\, (L+\HALF p h'(p)) e^{h(p)} ~.
$$ 
It follows that the solution of Eq.\equ(mmm) is
$$
U_{n,\tau} \,=\, e^{(e^{\tau} -1)X_n(pe^{-\tau/2})} e^{\tau} L  ~,
$$
as one can check by explicit computation.

From the explicit form of $X_n$, (in particular, the factor of $-n^4$),
and the estimates derived in \clm(semigroupa), we see that
for any $x_n \in H_{q,r}$, we have
$$
\| U_{n,\tau} x_n \|_{q,r} \,\le\, C \exp( - c_0 (e^\tau -1) n^4 ) 
e^{\tau q/2 } \|x_n\|_{q,r} ~.
$$
Combining this with the Remark of \equ(521), we immediately obtain
\CLAIM Lemma(ubound) If $U_\tau$ satisfies 
$$
e^{-\tau} \partial_\tau U_\tau  \,=\, M_{\exp(-\tau/2),2} U_\tau ~,
$$
with $U_0=1$, then there exist a $C(r,q,\nu) > 0$, and a $c_0>0$ such
that for any $w \in \Hqrn$,
$$
\|U_{\tau } w\|_{q,r,\nu}
\,\le\,C \exp(-e^{c_0 \tau /2}) \|w\|_{q,r,\nu}~.
\EQ(ubound)
$$

To complete the proof of \clm(semigroup2), we also need an estimate
of the semigroup generated by $M_{\eta,1}$. This is simply obtained,
however, because from \equ(m1def) we see that 
$M_{\eta,1} = M_{\exp(-\tau/2),2,1}$, restricted to functions whose
Fourier transform is supported away from the origin. Using this fact,
and the explicit formula given above for $M_{\exp(-\tau/2),2,n}$, we
see immediately that for any $w_1 \in H_{q,r}$, there exists
a constant $c_1 > 0$, such that if $U_{\tau,1}$ is the semigroup
generated by  $M_{\eta,1}$ one has
$$
\| U_{\tau,1} w_1\|_{q,r} \,\le\, C e^{-\c_1 \tau} \| w_1 \|_{q,r}~.
$$
\def\actualnumber{C} 
\SECTION The pseudo center manifold theorem for the singular system
Eq.\equ(full)

In this section, we prove \clm(full). Before we start with the proof,
we wish to point out in which sense we are here confronted with a new
problem, which does not allow for a straightforward application of
results from the literature. If we write the
system Eq.\equ(full) in the form 
$$
\eqalign{
\partial_\tau x_1 \,&=\, A_{1} x_1 + N_1( x_1,\eta, x_2 ,x_3)~,\cr
\partial_\tau \eta\,&=\, -\HALF\eta~,\cr
\partial_\tau x_2 \,&=\, A_{2} x_2 + N_2( x_1,\eta, x_2 ,x_3)~,\cr
\partial_\tau x_3 \,&=\, \eta^{-2}A_{3,\eta} x_3 
+ \eta^{-2}N_3( x_1,\eta, x_2 ,x_3)~,\cr
}
\EQ(full1)
$$
then, in view of the spectral properties of Eq.\equ(spectrum), there
is a ``gap'' between the ``central'' part (corresponding to $x_1$ and
$\eta $)
and the ``stable'' part (corresponding to $x_2$, $x_3$). The problem
is that we are really dealing with a singular perturbation because the
non-linearity in the equation for $x_3$ also diverges as $\eta
\downarrow 0$. This problem would be more easily overcome if $A_2$
were bounded. In that case, for sufficiently small $\eta$, the spectra
of $A_2$ and $\eta^{-2}A_3$ would not overlap, and we could define
first an invariant manifold by ``eliminating'' $x_3$, and then the
true invariant manifold by eliminating $x_2$ from the equations
obtained after elimination of $x_3$. However, since the spectra
overlap for all values of $\eta $, we resort to a strategy which
consists of a converging sequence of alternate eliminations of $x_2$
and $x_3$.

To define these successive eliminations,
we consider two equivalent representations of
Eq.\equ(full), one being Eq.\equ(full1) above and the other being
$$
\eqalign{
\partial_t x_1 \,&=\, \eta^2 \bigl (A_{1} x_1 + N_1( x_1,\eta, x_2
,x_3)\bigr )~,\cr
\partial_t \eta\,&=\, -\HALF\eta^3~,\cr
\partial_t x_2 \,&=\, \eta^2 \bigl (A_{2} x_2 + N_2( x_1,\eta, x_2
,x_3)\bigr )~,\cr
\partial_t x_3 \,&=\, A_{3,\eta} x_3 + N_3( x_1, \eta,x_2 ,x_3)~.\cr
}
\EQ(full2)
$$
We shall again omit the index $\eta$ from $A_3$.
We obtain Eq.\equ(full2) from Eq.\equ(full1) by rescaling the
evolution parameter of the autonomous system as $t+t_0 = \exp(\tau)$. 
(Note
that time is really given by $1/\eta^2-t_0$, while we view
$t$ and $\tau$ as the
evolution parameters of the vector fields.) 
We will call $\phicenter(\tau)$ the flow corresponding to Eq.\equ(full1)
and $\phistable(t)$ the flow corresponding to Eq.\equ(full2).
A simple inspection of the definition of these flows yields the useful
identity:
$$
\phicenter(\tau=\log {(y+t_0 )} )(\xi,x )\,=\,
\phistable(t=y )(\xi,x )~,
\EQ(ident)
$$
where 
$$\xi\,=\,(x_1,\eta),\quad\allx\,=\,(x_2,x_3)~.
\EQ(xix)
$$
We shall use the
relations \equ(xix) throughout. The identity \equ(ident) holds for
all $x_1$, $x_2$, $x_3$ and
for $\eta \ge0$. Note that the initial conditions are given for the
parameter $t=0$ and the parameter $\tau=\log(t_0)$, and that
$\eta(0)=t_0^{-1/2}$.
Thus, $\eta(0)$ is small if the parameter $t_0$ has been chosen
sufficiently large. (The bounds on the nonlinearities are uniform in
$t_0\ge t_0^*$ as follows from the calculations.)

Let $h_0$ be a function of $\xi$.
This function will always be an approximate invariant manifold for one
of two problems. To define these problems, we first introduce 
two effective non-linearities
$$
\eqalign{
F_j (h_0;\xi,x_2 )\,&=\, N_j\bigl (x_1,\eta,x_2,h_0(\xi ) \bigr
)~,\quad {\rm for~}j=1,2~,\cr
G_j  (h_0;\xi,x_3)\,&=\, N_j\bigl (x_1,\eta,h_0(\xi ),x_3 \bigr
)~,\quad {\rm for~}j=1,3~.\cr
}
$$
We then define two equations (corresponding to the two different
time scales Eq.\equ(full1) and Eq.\equ(full2)
of the same problem Eq.\equ(full)):
The first equation will be called the {\em center system}:
$$
\eqalign{
\partial_\tau x_1 \,&=\, A_{1} x_1 + F_1(h_0; \xi, x_2 )~,\cr
\partial_\tau \eta\,&=\, -\HALF\eta~,\cr
\partial_\tau x_2 \,&=\, A_{2} x_2 + F_2(h_0;\xi, x_2 )~.\cr
}
\EQ(center)
$$
Similarly, we define the
{\em stable system}
$$
\eqalign{
\partial_t  x_1 \,&=\, \eta^2 A_{1} x_1 + \eta^2 G_1(h_0; \xi,x_3)~,\cr
\partial_t  \eta\,&=\, -\HALF \eta^3~,\cr
\partial_t  x_3 \,&=\, A_{3} x_3 + G_3(h_0; \xi,x_3)~.\cr
}
\EQ(stable)
$$
Assume now that $h_2$ and $h_3$ are two given functions of $x_1$ and
$\eta$.
We define a map
$$
\FF~:~ \left ( {h_2 \atop h_3 } \right ) \mapsto \left ( {h_2' \atop
h_3'}\right )~,
$$
through the following construction:
We let
$h_2'(\xi )$ be the function whose graph is the
invariant manifold for the center system Eq.\equ(center) with
non-linearity $F_j(h_3;\xi,x_2 )$, 
and similarly we let
$h_3'(\xi )$ be the function whose graph is the invariant
manifold for the stable system Eq.\equ(stable) with non-linearity
$G_j(h_2;\xi,x_3 )$.
Our main result here is
\CLAIM Proposition(fff) The map $\FF$ has a fixed point $(h_2^*, h_3^*)$.
This fixed point provides an invariant manifold for the system
Eq.\equ(full).

\REMARK We shall in fact show that $\FF$ is a contraction in a
suitable function space. In particular, we show
that $\FF^n(0,0)$, the $n$-fold iterate
of $\FF$, converges to the limit $(h_2^*,h_3^*)$. The intuitive approach
behind this construction is that the $\FF^n(0,0)$ provide a sequence
of successive approximations to invariant manifolds
for the Eqs.\equ(stable) and \equ(center), in which the non-linearities
at the $n^{\rm th}$ step are given by the approximate solutions for
the invariant manifold problem of the other equation: The
non-linearities are then $F_j(\h(n-1,3); \xi,x_2 )$ (in
Eq.\equ(center)) and $G_j(\h(n-1,2); \xi,x_3 )$ (in
Eq.\equ(stable)).
\PROOF That the systems of equations \equ(center) and \equ(stable) have
invariant manifolds follows from our estimates (given
in Appendix B) on the semi-group generated by the linear operators $A_2$ and
$A_3$, and our estimates on the non-linear terms. (For expositions of
this theory that are particularly relevant in the present context, see
{\it e.g.}, [H, M, G].) The functions $h_2^*$
and $h_3^*$ whose graphs define the invariant manifolds satisfy
well known integral equations, see below.

Fix $h=(h_2,h_3)$ and consider the Eq.\equ(center). We want to find
the
function $h'_2(h;\xi)$ which eliminates $x_2$. 
To construct $h_2'$, we first consider the equation
$$
\eqalign{
\partial_\tau x_1\,&=\,A_1 x_1 + F_1\bigl
(h_3;\xi,h_2(\xi)\bigr )~,\cr
\partial_\tau \eta\,&=\,-\HALF\eta~.\cr
}
\EQ(centerflow)
$$
This is a differential equation on a finite dimensional space and we let
$\psic_\tau (\xi;h)$ denote the corresponding flow.
(Of course, the $\eta$-component of this problem can be explicitly
integrated.) 
We can then formulate the problem of finding the
invariant manifold which eliminates $x_2$ from Eq.\equ(stable) by
looking at the map defined by $h\mapsto \FF_2(h)$ where
$$
\FF_2(h)\,=\,
\int_{-\infty }^0 d\tau \,e^{-A_2\tau} F_2\bigl (h_3;\psic_\tau (\xi;h)
,h_2(\psic_\tau (\xi;h))\bigr )~.
\EQ(h2equ)
$$
(A particularly clear derivation of these equations can be found in
[G].) 
In a similar way, we define the flow $\psis_\tau (\xi;h)$ for the
equation
$$
\eqalign{
\partial_t x_1\,&=\,\eta^2 A_1 x_1 +\eta^2 G_1\bigl
(h_2;x_1,h_3(\xi)\bigr )~,\cr
\partial_t \eta\,&=\,-\HALF\eta^3~,\cr
}
\EQ(stableflow)
$$
and the map 
$$
\FF_3(h)\,=\,
\int_{-\infty }^0 dt \,e^{-A_3t} G_3\bigl (h_2;\psis_t (\xi;h)
,h_3(\psis_t (\xi;h))\bigr )~.
\EQ(h3equ)
$$
We now specify the function spaces in which we work. Recall
that $x_1\in \real^N$, $\eta\in \real$ and
that $\xi\in\real^{N+1}$. We let $\EE^\c = \real^N\oplus\real$
with the usual Euclidean norm. We also assume that $\EE^2$ and
$\EE^3$ are the Banach spaces in which the $x_2$ and $x_3$ live. In
our problem, these Banach spaces are the Hilbert spaces $H_{q,r}$ and
$\Hqrn$,
but
since we believe the present theory of singular vector fields may have
further applications, we consider the more general case for the
moment (see, for example,
[W2]). These Banach spaces should have the $\CC^k$ extension property [BF].
The functions $h_2$ and $h_3$ will be Lipshitz functions from
a ball of radius $r$ in $\EE^2$ and $\EE^3$, respectively. They
satisfy $h_j(0)=0$ and
are tangent at the origin to $\EE^j$, for $j=2,3$.
Thus, we define the metric spaces, for $j=2,3$:
$$
\HH_{j,\sigma} \,=\, \bigl \{
h_j : \EE^\c \to \EE^j ~\bigm |~
h_j(0)=0, \| h_j(\xi)-h_j(\tilde \xi)\|_{\EE^j}\,\le\, \sigma 
 \| \xi-\tilde \xi \|
\bigr \}~.
$$ 
We also define a distance
$$
\rho _{\HH_{j,\sigma}}(h_j,\tilde h_j) \,=\,\sup_{\xi\ne0}
{\|h_j(\xi)-\tilde h_j(\xi)\|_{\EE_j}\over \|\xi\|}~,
\EQ(bbb4)
$$
and introduce the notation
$$
\rho _{\HH_{\sigma}}(h,\tilde h) \,=\,\rho
_{\HH_{2,\sigma}}(h_2,\tilde h_2)
+\rho _{\HH_{3,\sigma}}(h_3,\tilde h_3)~.
$$
Standard results about the existence and uniqueness of solutions  of
systems of differential equations now imply that
$$
\bigl \|
\psic_\tau (\xi;h)-\psic_\tau (\tilde\xi;h)
\bigr \| \,\le\,
Ce^{\beta_2|\tau |} \|\xi-\tilde\xi\|~,
\EQ(bbb1)
$$
while
$$
\bigl \|
\psic_\tau (\xi;h)-\psic_\tau (\xi;\tilde h)
\bigr \| \,\le\,
Ce^{\beta_2|\tau |} \rho _{\HH_{\sigma}}(h,\tilde h)~,
\EQ(bbb2)
$$
for any $\beta_2>(N-1)/2$. Analogous estimates hold for the flow
$\psis$, though in that case one can choose any exponential growth rate
$\beta _3>0$, provided $|\eta |$ is sufficiently small. 
This is due to the presence of the factor of $\eta^2 A_1$ in the first
equation of Eq.\equ(stable).

With this in mind we define two more metric spaces (for $j=2,3$):
$$
\eqalign{
\KK_{j,\beta _j,D_j}\,=\,
\bigl\{
&\Psi_\tau : \real^+\times \EE^\c \times \HH_{2,\sigma} \times
\HH_{3,\sigma} \to 
\EE^\c~\bigm |\cr
&
\Psi_0(\xi;h)=\xi,
\Psi_\tau (0;h)=0, \Psi_\tau  {\rm~is~} \CC^1 {\rm ~in~}\tau ,\cr
&\| \Psi_\tau (\xi,h)-\Psi_\tau(\tilde \xi,\tilde h)\|
\,\cr &~~~~~~~~~~~\,\le\,D_j e^{\beta _j|\tau |}
\bigl (
\|\xi-\tilde \xi\|+ \rho _{\HH_{\sigma}}(h,\tilde h)\|\xi\|
\bigr ) 
\bigr \}~,\cr
}
$$
with a corresponding Lipshitz metric 
$$
d_j(\Psi,\tilde \Psi)\,=\,
\|\Psi-\tilde \Psi\|_{\KK_j}~,
\EQ(bbb3)
$$
where 
$$
\|\Psi\|_{\KK_j}\,=\,\sup _{t\ge0} \sup_ {\xi\in \EE^\c\atop \xi\ne0} {
e^{\beta _jt}\|\Psi_t(\xi)\|\over \|\xi\|}~.
$$
These spaces are modeled on those used in [EW].
\REMARK Since we are interested in {\em local} invariant manifolds,
we will assume that the non-linear terms have been cut off outside a
ball of radius $r$ in each of their arguments. Since in the
applications of this paper all our functions are elements of
Hilbert spaces, we can assume that there exist smooth cut-off
functions which are equal to 1 inside a ball of radius $r/2$ and are
equal to zero outside a ball of radius $r$, and we multiply each of
the non-linear terms in Eq.\equ(full) by such a cutoff. For example, in
Eq.\equ(stable), we certainly need to cutoff the function $\eta ^2$ by
$\eta^2\chi(\eta)$ (where $\chi$ is the cutoff function) to avoid
blowup problems.

Given this setup, we show that the map $\FF$ is a contraction of
$\HH_{2,\sigma}\times \HH_{3,\sigma}$.
In terms of the notation given above $\FF$ is now defined as
$\FF(h)=\bigl (\FF_2(h),\FF_3(h)\bigr )$.
One must first show that $\FF$ maps
this space to itself. This step is however an easy variant of the
argument which shows that $\FF$ is a contraction, and we leave it as
an exercise to the reader. To show that $\FF$ is a contraction, we use
the maps \equ(h2equ) and \equ(h3equ). Then we
see that the ``$j$'' component, $j=2,3$,
of $\FF(h_2,h_3)(\xi)-\FF(\tilde h_2,\tilde
h_3)(\xi)$ is given by
$$
\eqalign{
\Delta_j\,=\,
\int_{-\infty }^0 d\tau \,e^{-A_j \tau }\biggl (
&U_j(h;\xi,\tau)
-U_j(\tilde h;\xi,\tau)\biggr )~,\cr
}
\EQ(Fdiff)
$$
where
$$
\eqalign{
U_2(h;\xi,\tau)\,&=\,F_2\bigl (h_3;\psic_\tau (\xi;h),h_2\bigl (\psic_\tau
(\xi;h)\bigr )\bigr )\cr\,&=\,N_2\bigl (\psic_\tau (\xi;h),h_2\bigl (\psic_\tau
(\xi;h)\bigr ),h_3\bigl (\psic_\tau
(\xi;h)\bigr )\bigr )~,\cr
U_3(h;\xi,\tau)\,&=\,G_3\bigl (h_2;\psis_\tau (\xi;h),h_3\bigl (\psis_\tau
(\xi;h)\bigr )\bigr )\cr\,&=\,N_3\bigl (\psis_\tau
(\xi;h),h_2\bigl (\psis_\tau
(\xi;h)\bigr ),h_3\bigl (\psis_\tau
(\xi;h)\bigr )\bigr )~,\cr
}
$$
{\it cf.}~Eqs.\equ(center), \equ(stable).
Consider now $\Delta_2$. From the estimates on the non-linear term
$N_2$ in Eq.\equ(full), we see that $\FF_2$ is a multi-linear function
of its arguments. Thus, we can estimate the difference in the
integrand of $\Delta_2$ by the sum of the differences in the arguments
of $\FF_2$, multiplied by the Lipshitz constant of $\FF_2$. Because we
have cutoff $\FF_2$ outside a ball of radius $r$, this Lipshitz
constant can be made arbitrarily small by making $r$ sufficiently
small. Thus, calling this Lipshitz constant $\ell_2(r)$, we see from
the estimates on $e^{A_2t}$ which follow from the results of Appendix B
and from
Eqs.\equ(bbb4)--\equ(bbb3) that
$$
\eqalign{
\|\Delta_2\|_{\EE^2} \,&\le\,
\int_0^\infty d\tau\, {C\over \sqrt{\tau }} e^{-N\tau /2} \ell_2(r)
\biggl (
\rho_{\HH_\sigma}(h,\tilde h)\cr
&+\|\psic_\tau (\xi,h)-\psic_\tau
(\xi,\tilde h)\| + \|h_2\bigl (
\psic_\tau (\xi,h)\bigr )-\tilde h_2\bigl (\psic_\tau (\xi,\tilde h)\bigr )\|_{\EE^2} 
\biggr )\cr
\,&\le\,
\int_0^\infty d\tau\,{C\over \sqrt{\tau }} e^{-N\tau /2} \ell_2(r)
\biggl (\rho_{\HH_\sigma}(h,\tilde h)\cr
&
+\rho_{\HH_\sigma}(h,\tilde h) C e^{\beta _2\tau }+
\rho_{\HH_\sigma}(h,\tilde h) C e^{\beta _2\tau }\biggr )\,\le\,\const
\ell_2(r) \rho_{\HH_\sigma}(h,\tilde h) ~.
}
$$
Thus, we have shown that $\FF$ is a contraction.

We next consider the manifold $\MM$ given by $(\xi,h_2^*(\xi),
h_3^*(\xi))$---where $x_1$ is in a small neighborhood of 0
and $\eta$ is in a small positive interval $0\le \eta \le \eta
_0$. We want to show that $\MM$
is indeed an invariant manifold for the full system
Eq.\equ(full). From this it follows,
since the flows $\phis$ and
$\phic$ are equivalent, up to rescaling of time,
that $\MM$ is also an invariant manifold
for the Eqs.\equ(full1)
and \equ(full2). If we set $x_2=h_2^*(\xi )$ and
$x_3=h_3^*(\xi )$, then the third equation of Eq.\equ(full2) is
satisfied because the third equation, when restricted to the manifold
$x_2=h_2^*$ is just the second equation of the stable system
Eq.\equ(stable). with non-linearity $G_3(h_2^*; \dots)$. To see that
the remaining equations are satisfied just note that the 
first, second and fourth equations in the full system Eq.\equ(full)
become, after rescaling of time,
$$
\eqalign{
\dot x_1 \,&=\, A_1 x_1 + N_1(x_1,\eta,x_2,x_3 )~,\cr
\dot \eta \,&=\,-\HALF\eta ^3~,\cr
\dot x_2 \,&=\, A_2 x_2 + N_2(x_1,\eta,x_2,x_3 )~,\cr
}
$$
and if we set $x_2=h_2^*$ and $x_3=h_3^*$, we see that we are just on
the invariant manifold for the center system Eq.\equ(center). Hence,
we have found the invariant manifold for the full system
Eq.\equ(full). 
\def\actualnumber{D}
\SECTION The vanishing of the non-linearity at zero
momentum

In this appendix, we prove \clm(x3). This proof
is essentially a scaling
argument. We shall study the nonlinearity
$N_1(x_1,\eta,x_2,x_3)$ and we restrict it to the invariant
manifold, {\it i.e.}, we replace it by $\tilde N_1(x_1,\eta)$ and let
$\eta$ go to 0. In particular, we shall show that only one term
survives, namely the one which is cubic in $x_1^3$, and all others go
to 0 as $\eta\to0$.

To prove this,
we will analyze the nonlinearities $N_j$ term by term, using their
definitions as given in Eqs.\equ(wequ2) and \equ(full). Recall again
that $A_1=0$ since we are considering here the projection onto the
first eigenvalue of $L$. In Eq.\equ(wequ2), the nonlinearities are
given by the terms $f_2$, $f_3$, $f_4$, and $g$, and these have been
bounded in \clm(nonlin) and \clm(nonlin2). Recall finally that every
factor of $w^\c $ contributes a factor of $e^{-\tau /2}=\eta$ and
every factor of $w^\s$ contributes a factor of $e^{-\tau }=\eta^2$ to
these bounds. 

Begin by considering the contribution from $f_2$. According to
Eq.\equ(t1), we can extract {\em another factor} of $\eta$ from
Eq.\equ(t111), by using the quadratic nature of $K_2$, {\it cf.}~\clm(k2k3). 

Using \clm(nonlin) and  \clm(nonlin2), we see that the only 
contributions from $f_3$, $f_4$, and
$g$ which do not vanish as $\eta\to0$ are those of the type $(w^\c)^3$
in $f_3$, of the type $w^\s
(w^\c)^2$ in $f_4$, and of the type $(w^\c)^2$ in $g$.

We start by analyzing $f_3$. If we write it out,
we find
$$
\eqalign{
\eta^{-2}\bigl (f_3(w^\c)\bigr )(p)\,&=\,\eta^{-2}\chi\bigl (\Phi(p\eta)\bigr )
\int dx \,  \bar \phi_{\Phi(p\eta)}(x)\cr 
&\times\eta^2\int_{\eta^{-1}\Phi(-1/2)}^{\eta^{-1}\Phi(1/2)}\kern -2em dp_1\,dp_2\,
\Phi'(p_1\eta)\Phi'(p_2\eta) \cr&\times
\phi_{\Phi(p_1\eta)}(x)\,\phi_{\Phi(p_2\eta)}(x)\,\phi_{\Phi(p\eta)-\Phi(p_1\eta)
-\Phi(p_2\eta)}(x)\cr
&\times
w^\c(p_1)\,w^\c(p_2)\,w^\c\bigl (\eta^{-1}\Phi^{-1}(\Phi(p\eta)-\Phi(p_1\eta)
-\Phi(p_2\eta))\bigr )~,\cr
}
$$
{\it cf.}~Eq.\equ(x00). Upon taking $\eta\to0$, this converges to
$$
\chi(0)\int dx\,\bar\phi_0(x) \phi_0^3(x) \int dp_1 \,dp_2\,w^\c(p_1)
w^\c(p_2) w^\c(p-p_1-p_2)~.
\EQ(ggg4)
$$

Analogous arguments can be used to discuss the ``surviving'' terms of
$f_4$ and $g$. We just summarize the steps analogous to the
calculation of $f_3$. One gets, as $\eta\to 0$,
$$
\eta^{-2} \bigl (f_4(w^\c,w^\s\eta,\eta)\bigr)(p)~\to~ 6\chi(0) \int dx\, \bar\phi_0(x) u_\epsilon (x)
\phi_0(x)
\int dp' w^\c(p') w^\s(p-p';x)~,
\EQ(ggg5)
$$
and
$$
\eta^{-1} \bigl (g(w^\c,w^\s\eta,\eta)\bigr)(p)~\to~ 
-3 u_\epsilon (x)\phi_0^2(x) \int dp' \,w^\c(p-p') w^\c(p')~.
\EQ(ggg6)
$$

We next study these limiting expressions in the basis
$\{\psi_n(p)\}_{n=0}^{\infty }$ of eigenfunctions of $L=-p^2-\HALF
p\partial_p$.
Then we can write $w^\c(p)$ as
$$
x_1 \psi_0(p)+\sum_{n=1}^\infty  x_2^{(n)} \psi_n(p)~.
\EQ(wcp)
$$
The crucial remark is now that {\em on the invariant manifold,
$x_2^{(n)}$ will be replaced by $h_2^{*,(n)}$}, and similarly $w^\s$
will be equal to $h_3^*$. We now compute the limiting forms of $h_2^*$
and $h_3^*$, and then we substitute these values in Eqs.\equ(ggg4)--\equ(ggg6).
Consider the equation for $h_3^*$.
Then from Eq.\equ(stableflow), we have
$$
\eqalign{
\partial_t x_1 \,&=\, \eta^2 G_1(h_2^*; x_1,h_3^*(\xi))~,\cr
\partial_t\eta\,&=\,-\HALF\eta^3~,\cr
}
$$
because we are considering the case $N=1$ where the linear part
vanishes.
We also have from Eq.\equ(h3equ),
$$
h_3^*(x_1,\eta)\,=\,
\int_{-\infty }^0 dt \,e^{-A_{3,\eta}t} G_3\bigl (h_2^*;\psis_t (x_1,\eta;h^*)
,h_3^*(\psis_t (x_1,\eta;h^*))\bigr )~.
\EQ(h3equ2)
$$
Now, when $\eta=0$, we have $\psis_t(\xi;h)=\psis_t(x_1,0;h)=x_1$, and
this reduces to
$$
\eqalign{
h_3^*(x_1,0)\,&=\,
\int_{-\infty }^0 dt \,e^{-A_{3,0}t} G_3\bigl (h_2^*;x_1,0
,h_3^*(x_1,0)\bigr )\cr
\,&=\,-A_{3, 0}^{-1}G_3\bigl (h_2^*;x_1,0
,h_3^*(x_1,0)\bigr ) ~.
}\EQ(h3equ2b)
$$
Note next that for $\eta=0$ we have $A_{3,0}=M_0$, {\it cf.}~Eq.\equ(full), and
this means $A_{3,0}=Q_{\rm per}L_{\rm per}$.
We denote by $\xi_n(x)$ the eigenfunctions and by $\sigma_n$ the
eigenvalues of $Q_{\rm per}L_{\rm per}$. Using Eq.\equ(loperator) and 
\clm(Eckhaus), we see that $\sigma_n=\lambda_{\ell=0,n-1}$ and therefore
they are
given by $\sigma_1=-\OO(\epsilon ^2)$ and
$\sigma_n\approx -(1-(n-1)^2)^2$, when $n\ne1$.
Then the $n^{\rm th}$ component (in this basis) of $h_3^*$ (at
$\eta=0$) is given by 
$$
h_3^{*,(n)}(p)	\,=\,
-\sigma_n^{-1} \cdot \bigl (-3 \int dx\,
\bar\xi_n(x) u_\epsilon (x) \phi_0^2(x)\bigr )
\int dp' w^\c(p-p') w^\c(p')~,
\EQ(h3star)
$$
since all other terms vanish in the limit $\eta\to0$.
We next substitute the value Eq.\equ(wcp) for $w^\c$ and
set $x_2 = h_2^*$ in Eq.\equ(h3star), and get
$$
\eqalign{
h_3^{*,(n)}(p)\,&=\,-{x_1^2\sigma_n^{-1}} \cdot\bigl (-3 \int dx\,
\bar\xi_n(x) u_\epsilon (x) \phi_0^2(x)\bigr )\cr
&\times
\biggl (\int dp'\, \psi_0(p')\,\psi_0(p-p') + \OO(x_1 h_2^* +
(h_2^*)^2)\biggr )~.\cr
}
$$
Next, we replace $w^\s$ in Eq.\equ(ggg5) with $h_3^{*}$, and 
in that same equation make the substitution for $w^{\c}$ that
we used above, and we find: 
$$
\eqalign{
18 x_1^3 \sum_{n=0}^\infty & \sigma_n^{-1} 
\bigl ( \int dx\,
\bar\xi_n(x) u_\epsilon (x) \phi_0^2(x)\bigr )
\bigl (\int dx' \,\bar \phi_0(x') u_\epsilon (x') \xi_n(x')\bigr )\cr
&\times \biggl (\int dp_1\, dp_2 \,
\psi_0(p_1 )\psi_0(p_2) \psi_0(p-p_1-p_2) + \OO(x_1 h_2^* +
(h_2^*)^2)\biggr )~.
\cr
}
\EQ(ggg10)
$$
Thus we see that the only terms which survive in $N_1$ and $N_2$ in
the limit $\eta\to0$ result from adding together Eqs.\equ(ggg10) and
\equ(ggg4). 
We obtain
$$
\eqalign{
X\,&=\,x_1^3\biggl \{
\int dx\,\bar\phi_0(x) \phi_0^3 (x) 
+
18\sum_{n=0}^\infty  \sigma_n^{-1}
( \int dx'\,
\bar\xi_n(x') u_\epsilon (x') \phi_0^2(x')\bigr )\cr
&\times\bigl (\int dx'' \,\bar \phi_0(x'')\, u_\epsilon (x'')\,
\xi_n(x'')\bigr )
\biggr \}\cr
&\times
\biggl (\int dp_1\, dp_2 \,
\psi_0(p_1)\, \psi_0(p_2)\, \psi_0(p-p_1-p_2) \biggr )~.\cr
}
\EQ(ggg11)
$$
This coefficient will turn out to be
exactly the same as that which appears below as the
coefficient of the cubic terms in the center manifold in the periodic
case, and since we know that in periodic case this coefficient (and indeed,
the entire nonlinear term) is zero, it must vanish in the present
case as well. The only remaining point in the proof of \clm(x3) is
the computation of the coefficient of the cubic term in the equation
in the center manifold in the periodic case, and we do that in
the following subsection.

\REMARK The above argument might seem incomplete since it ignores the 
${\OO}(x_1 h_2^* + (h_2^*)^2)$ error terms in \equ(ggg10). In fact,
those terms vanish for $x_1$ small. To see why, note that our computations of
the $\eta \to 0$ limit of $f_2$, $f_3$, $f_4$ and $g$ apply
also to the nonlinear term $N_2(x_1,\eta,h_2^*(\xi),h_3^*(\xi))$
in the equation for $h_2^*$ in \equ(IM). Thus, in the $\eta \to 0$ limit
$h_2^*$ satisfies:
$$
\partial_{x_1} h_2^*(x_1,0) \tilde{N}_1(x_1,0) \,=\,
A_2 h_2^*(x_1,0) + N_2(x_1,0,h_2^*(x_1,0),h_3^*(x_1,0))~.
$$
Using the estimates on $h_2^*$ and $h_3^*$ derived above,
we see that this equation implies $h_2^*(x_1,0) = 0$ for
all $x_1$ sufficiently small, and hence the error terms in \equ(ggg10) 
vanish.

\SUBSECTION The non-linearity in the periodic case

In this subsection we compute the explicit form
of the non-linearity (which we know to be 0 because the invariant
manifold is made up of fixed points in this case). But this explicit
form will allow us to compare it with the expression obtained in
Eq.\equ(ggg11) so that the proof of \clm(x3) will be complete.

We start from the equation
$$
\partial_\tau  v \,=\, L_{\rm per} v - 3 u_\epsilon  v^2  - v^3~.
\EQ(fax1)
$$
Let $y_0$ be the component of $v$ in the direction of the highest
eigenvalue, $\sigma_0=0$, of $L_{\rm per}$, and $y_n$, the projection onto the
directions $\xi_n$, defined after Eq.\equ(h3equ2b), associated to the
eigenvalues $\sigma_n$. Then the invariant
manifold can be written in the form
$$
y_n\,=\, Y_n(y_0)~,\quad n\,=\,1,2,\dots~.
\EQ(fax3)
$$
Using the fact that the eigenfunction with eigenvalue $0$
is $u'_{\epsilon}$, we can decompose $v$ as:
$$
v(x)\,=\, y_0 u_\epsilon '(x) +\sum_{n=1}^\infty  \xi_n (x)
Y_n(y_0)~,
\EQ(fax4)
$$
the projection of Eq.\equ(fax1) onto the invariant manifold leads to
$$
\partial_\tau y_0\,=\, - \int dx\, u_\epsilon '(x)\, \bigl (3 u_\epsilon
(x)v(x)^2+v(x)^3\bigr )~.
\EQ(fax5)
$$
Note that there is no linear term because $\sigma_0=0$.

We are interested in the exact form of the cubic term in $y_0$ on the
r.h.s.~of Eq.\equ(fax5). There
are two contributions, one from $v^3$, leading to
$$
-y_0^3\int dx\, u_\epsilon '(x)^4~,
\EQ(fax7)
$$ 
and one from the quadratic non-linearity:
$$
Y\,=\,-6 y_0 \sum_{n=1}^\infty Y_n^{(2)}(y_0) \int dx\, u_\epsilon '(x)\, u_\epsilon (x)\,
u_\epsilon '(x)\, \xi_n(x)~.
\EQ(fax8)
$$
Here, $ Y_n^{(2)}(y_0)$ is the quadratic term in $y_0$ of $Y_n$.
Substituting Eq.\equ(fax5) into the equation for $Y_n$, we find the
perturbative result:
$$
Y_n^{(2)}(y_0)\,=\,y_0^2\cdot 3\sigma_n^{-1}
\int dx\, \bar\xi_n(x)\, u_\epsilon (x)\, u_\epsilon '(x)^2~.
$$
Inserting into Eq.\equ(fax8), it is seen to become
$$
Y\,=\,-y_0^3 18\sum_{n=1}^\infty \sigma_n^{-1}\int dx\, u_\epsilon
 '(x)^2\,  u_\epsilon (x)\,
 \xi_n(x)\int dx'\, \bar\xi_n(x')\, u_\epsilon (x')\, u_\epsilon '(x')~.
\EQ(fax12)
$$
Combining Eqs.\equ(fax7) and \equ(fax12), we get the desired result,
namely that the cubic non-linearity in the periodic case coincides with the
quantity
$X$ of Eq.\equ(ggg11), provided we recall
that $\phi_0 = u_{\epsilon}'$. This completes the proof of \clm(x3).

\SECTIONNONR Acknowledgments 

This work was begun while J-P E was a
visitor at the Pennsylvania State University.  It was completed
during a visit of CEW to the University of Geneva.  The support
of the Shapiro Visitors Fund at Penn State and the hospitality of
the Department of Theoretical Physics at the University of Geneva
are gratefully acknowledged.  In addition, the authors' research is
supported in part by the Fonds National Suisse and the National
Science Foundation Grant DMS-9501225.

\SECTIONNONR References

\eightpoint
\raggedright
\widestlabel{[BKL]}
\ref 
 \no BF
 \by Bonic, R. and J. Frampton
 \paper Smooth functions on Banach manifolds
 \jour J. Math. Mech.
 \vol 15
 \pages 877--898
 \yr 1966
\endref
\ref
 \no BK
 \by Bricmont, J. and A. Kupiainen
 \paper Stable Non-Gaussian Diffusive Profiles
 \jour Nonlinear Analysis
 \vol 26
 \pages 583-593
 \yr 1995
\endref
\ref
 \no BKL
 \by Bricmont, J., A. Kupiainen, and G. Lin
 \paper Renormalization Group and Asymptotics of Solutions
 of Nonlinear Parabolic Equations
 \jour Comm. Pure Appl. Math.
 \vol 47
 \pages 893-922
 \yr 1994
\endref
\ref
\no C
\by Carr, J.
\book {The Centre Manifold Theorem and its Applications}
\publisher  New York, Springer-Verlag
\yr 1983
\endref
\ref
\no CE
  \by Collet, P. and J.-P. Eckmann
  \book Instabilities and Fronts in Extended Systems
  \publisher Princeton University Press
  \yr 1990
\endref
\ref
  \no CEE
  \by Collet, P., J.-P. Eckmann, and H. Epstein
  \paper Diffusive repair for the Ginzburg-Landau equation
  \jour  Helv. Phys. Acta
  \vol 65
  \pages 56--92 
  \yr 1992
\endref
\ref
  \no CR
  \by Crandall, M. and P. Rabinowitz
  \paper Bifurcation from simple eigenvalues
  \jour J. Funct. Analysis
  \vol 8
  \pages 321--340
  \yr 1971
\endref
\ref
\no E
  \by Eckhaus, W.
  \book Studies in non-linear stability theory
  \jour Springer tracts in Nat. Phil.
  \vol 6
  \publisher Berlin, Heidelberg, New York, Springer
  \yr 1965
\endref
\ref
\no EW
\by Eckmann, J.-P. and C. E. Wayne
\paper Propagating Fronts and the Center Manifold Theorem
\jour Comm. Math. Phys.
\vol 136
\pages 285-307
\yr 1991
\endref
\ref
\no G
  \by Gallay, Th.
  \paper A center-stable manifold theorem for differential equations in Banach spaces
  \jour Commun. Math. Phys.
  \pages 249--268
  \vol 152
  \yr 1993 
\endref
\ref 
\no GJ
\by Glimm, J. and Jaffe A.
\book Quantum Physics, A Functional Integral Point of View
\publisher New York, Springer-Verlag
\yr 1981
\endref
\ref
\no H
\by Henry, D.
\book  Geometric Theory of Semilinear Parabolic Equations
\jour Lecture Notes in Mathematics
\vol 840
\publisher New York, Springer-Verlag
\yr 1981
\endref
\ref
\no L
  \by Levine, H.A.
  \paper The role of critical exponents in blow up theorems
  \jour SIAM Review
  \vol 32
  \pages 262-288
   \yr 1990
\endref
\ref 
\no M
\by Mielke, A.
\paper A new approach  to sideband-instabilities using the principle
  of reduced instability
\inbook  Nonlinear Dynamics and Pattern Formation in the Natural
Environment
\bybook Doelman, A., van Harten, A., eds.
\publisher Longman, UK,
\yr 1995
\pages 206--222
\endref
\ref
\no Sch
\by Schneider, G.
\paper Diffusive stability of spatial periodic solutions of the
Swift-Hohenberg equation
\jour Commun. Math. Phys.
\vol 178
\pages 679--702
\yr 1996
\endref
\ref
\no W
\by Wayne, C.E.
\paper Invariant manifolds for parabolic partial differential
equations on unbounded domains
\jour Arch. Rat. Mech.
\toappear
\endref
\ref
\no W2
\by Wayne, C.E.
\paper Invariant manifolds and the asymptotics of parabolic
equations in cylindrical domains
\jour Proceedings of the China/US conference on differential equations
and applications, Hangzhou, PRC July 1996.
\toappear
\endref
}
\bye